\documentclass[fleqn,usenatbib]{mnras}
\usepackage{newtxtext,newtxmath}
\usepackage{colortbl}

\usepackage[T1]{fontenc}

\DeclareRobustCommand{\VAN}[3]{#2}
\let\VANthebibliography\thebibliography
\def\thebibliography{\DeclareRobustCommand{\VAN}[3]{##3}\VANthebibliography}

\usepackage{graphicx}	% Including figure files
\usepackage{amsmath}	% Advanced maths commands
\usepackage{mathtools}
\usepackage{xcolor}
\usepackage{soul}
\title[Lens detection with the ILT]{A machine learning based approach to gravitational lens identification with the International LOFAR Telescope}

\author[S. Rezaei et. al]{
S. Rezaei,$^{1,2,3}$\thanks{E-mail: rezaei@astro.rug.nl} 
J. P. McKean,$^{1,3}$ 
M. Biehl,$^{2,4}$ 
W. de Roo$^{1}$ and
A. Lafontaine$^{1}$
\\
$^{1}$Kapteyn Astronomical Institute, University of Groningen, Postbus 800, NL-9700 AV Groningen, The Netherlands\\
$^{2}$Bernoulli Institute for Mathematics, Computer Science and Artificial Intelligence, University of Groningen, Postbus 407, NL-9700 AK Groningen, The Netherlands\\
$^{3}$ASTRON, Institute for Radio Astronomy, Oude Hoogeveensedijk 4, 7991 PD Dwingeloo, The Netherlands\\
$^{4}$SMQB, Institute of Metabolism and Systems Research IBR Tower, Level 3 College of Medical and Dental Sciences, University of Birmingham, Birmingham, UK\\
}

\date{Accepted XXX. Received 2022 July 20; in original form 2022 June 11}

\pubyear{2022}

\begin{document}
\label{firstpage}
\pagerange{\pageref{firstpage}--\pageref{lastpage}}
\maketitle

\begin{abstract}
We present a novel machine learning based approach for detecting galaxy-scale gravitational lenses from interferometric data, specifically those taken with the International LOFAR Telescope (ILT), which is observing the northern radio sky at a frequency of 150~MHz, an angular resolution of 350~mas and a sensitivity of 90~ \textmu Jy~beam$^{-1}$ ($1\sigma$). We develop and test several Convolutional Neural Networks to determine the probability and uncertainty of a given sample being classified as a lensed or non-lensed event. By training and testing on a simulated interferometric imaging data set that includes realistic lensed and non-lensed radio sources, we find that it is possible to recover 95.3 per cent of the lensed samples (true positive rate), with a contamination of just 0.008 per cent from non-lensed samples (false positive rate). Taking the expected lensing probability into account results in a predicted sample purity for lensed events of 92.2 per cent. We find that the network structure is most robust when the maximum image separation between the lensed images is $\geq3$ times the synthesized beam size, and the lensed images have a total flux density that is equivalent to at least a $20\sigma$ (point-source) detection. For the ILT, this corresponds to a lens sample with Einstein radii $\geq0.5$~arcsec and a radio source population with 150 MHz flux densities $\geq2$~mJy. By applying these criteria and our lens detection algorithm we expect to discover the vast majority of galaxy-scale gravitational lens systems contained within the LOFAR Two Metre Sky Survey.
\end{abstract}


\section{Introduction}

Strong gravitational lensing occurs when the light from a distant galaxy is deflected due to the space-time curvature that is caused by another galaxy along the line of sight. As a result of this  phenomenon, the foreground galaxy acts like a lens, and can produce multiple magnified images of the background galaxy (see \citealt{Treu2010} for a review). 

Ever since the first gravitational lens was discovered by \citet*{Walsh1979}, gravitational lensing has become a powerful tool to test models for galaxy formation and cosmology. For example, over the last four decades, gravitational lensing has been used to  measure the mass components of massive early-type galaxies \citep{Treu2006,Bolton2008,Auger2009,Auger2010}, constrain their stellar initial mass function \citep{Spiniello2012,Spiniello2014}, and determine their inner mass density profiles \citep{Wucknitz2004,Koopmans2006,Spingola2018}. Also, through detailed modelling of the surface brightness distribution of the lensed images, it has been possible to place constraints on the nature of dark matter \citep{Vegetti2012,Vegetti2014,Ritondale2019,Hsueh2020,Gilman2020}. When such modelling is combined with a time-delay observed between the different lensed images of a flux-variable background object, models for the expansion of the Universe and dark energy have also been tested \citep{Suyu2010,Suyu2013,Bonvin2017,Wong2020}.

 There have been several dedicated surveys to find gravitational lenses across the entire electromagnetic spectrum with both imaging and spectroscopic data (e.g. \citealt{Patnaik1992,Myers2003,Bolton2006,Negrello2010,More2012,Treu2018,Spiniello2018}). These surveys have discovered hundreds of examples of strong gravitational lensing, where the background galaxy is either a compact source associated with an Active Galactic Nucleus (AGN; quasar) or an extended source, primarily associated with the emission from stars in a galaxy (e.g. \citealt{King1999,Browne2003,Bolton2008,Faure2008,Wardlow2013,Negrello2017,Lemon2018,Lemon2019,Lemon2020,Li2021}). 
 
Unfortunately, gravitational lensing by massive galaxies is quite a rare event, with one gravitational lens found in about a thousand galaxies observed \citep{Chae2002,Wardlow2013,Amante2020}. This makes their identification from visual inspection both time consuming and prone to incompleteness (e.g. \citealt{Jackson2008,Marshall2016,More2016}), as the parent population that needs inspecting tends to be of order $10^4$ galaxies. Therefore, the vast majority of the gravitational lenses discovered thus far have been found through applying a set of selection criteria in catalogue space, based on the optical colour or radio spectral index, the total flux density and the morphology of the candidate lensed images. However, some level of visual inspection is still needed to verify potential lens candidates.

In the near future, the ever-increasing size of datasets from existing and proposed wide-field surveys necessitates sophisticated automated search techniques to identify new lens candidates. This is because with parent samples of order $>10^7$ galaxies, even applying various selection criteria will still result in a prohibitively large number of candidates requiring visual inspection. For example, it is expected from the large-scale imaging surveys to be carried out with the Vera C. Rubin Observatory, the {\it Nancy Grace Roman Space Telescope} and {\it Euclid} at optical/infrared wavelengths, and with the next generation Very Large Array (ngVLA) and the Square Kilometre Array (SKA) at radio wavelengths, that more than $10^5$ gravitationally lensed galaxies will be discovered \citep{Oguri2010, Collett2015, McKean2015}. 

To test various identification techniques, \citet{Metcalf2019} recently carried out a lens finding challenge that focused on optical/infrared datasets. They tested a wide variety of automated techniques developed by the community, such as gravitational arc and ring finders \citep{Cabanac2007,Sonnenfeld2018}, machine learning and deep learning algorithms \citep{Hartley2017,Petrillo2017,Schaefer2018,Lanusse2018,Avestruz2019}, and also brute force visual inspection methods \citep{Jackson2008}. They found that more than 50 per cent of lens systems could be identified, without any false positive events, using the automated approaches when  certain thresholds on the lensed image brightness or size were applied. Significantly, the automated methods outperformed the brute-force visual inspection by experts in the field.

Arc and ring finder algorithms do not use any learning techniques \citep{Cabanac2007,Sonnenfeld2018}, but instead fit parametric lens models to any detected arc-like surface brightness distribution to determine the likelihood of it being lensed. The need to fit a model to each detected arc or ring can become computationally expensive, and the performance of such algorithms is limited to what we expect a lens system to look like. Moreover, the risk of a false positive detection for cases where spiral arms or tangentially elongated star forming regions appear as arc-like features can be a problem for these types of algorithms. On the other hand, learning based techniques, in particular Convolutional Neural Networks (CNNs; \citealt{Hezaveh2017,Petrillo2017,Petrillo2019, Jacobs2017,Jacobs2019b,Jacobs2019a,Cheng2020,Akhazhanov2021,Gentile2022}) and Support Vector Machines (SVMs; \citealt{Hartley2017}), have been recently developed for detecting and modelling strong gravitational lens systems. 

The advantage of using deep learning algorithms over traditional lens finding algorithms are multifold. They are less computationally expensive and can reduce the need for user involvement. The model-agnostic nature of learning algorithms is free to capture underlying structure that model-dependent methods may miss. However, it is still needed for users to provide the training data and label them into classes of lensed and non-lensed, leaving the machine to learn by itself the important features that describe a gravitational lens system. In this regard, deep learning is well-suited to identifying lensed features, as their surface brightness distribution is highly correlated via the lens equation. Indeed, recent applications of deep learning to lens identification in wide-field ground-based optical/infrared surveys have found hundreds to thousands of gravitational lens candidates \citep{Li2021,Rojas2021}. Although the vast majority of these candidates have still to be confirmed as genuine gravitational lenses, this highlights how powerful such techniques can be.

To date, there has been almost no research done in applying machine learning techniques for detecting gravitational lenses with radio interferometers; although, see \citet{Morningstar2018, Morningstar2019} for a discussion on using deep learning for image deconvolution and lens modelling. This is in part due to the availability of large amounts of wide-field multi-band optical imaging data from the Kilo Degree Survey (KiDS) and the Dark Energy Survey (DES), and the impending launch of {\it Euclid} (currently planned for 2023). Also, as the data from interferometers is in the native visibility plane, this apparent complexity has made such studies seem additionally challenging. However, the next generation of radio interferometers will have the sensitivity and angular resolution to be excellent gravitational lens finding machines, with several unique science applications (e.g. \citealt{McKean2015}).

Here, we focus on developing a lens detection algorithm that can be applied to large-area surveys (15\,000~deg$^2$) at high angular resolution (5 to 500 mas synthesized beam size) with next generation radio interferometers. In particular, we concentrate on lens surveys with the Low Frequency Array (LOFAR; \citealt{vanHaarlem2013}), which mainly operates between 120 and 170 MHz, and has the sensitivity to detect the non-thermal emission from lensed radio sources \citep{Stacey2019} and the long baselines needed to resolve their structure \citep{Badole2021}. Also, given that the International LOFAR Telescope (ILT) has now gone through the commissioning phase \citep{Morabito2021,Jackson2021,Bonnassieux2021,Harwood2021,Sweijen2021,Timmerman2021}, the routine imaging of the survey data with the international stations of LOFAR will soon start \citep{Sweijen2022}. Therefore, developing methods for identifying lensed radio galaxies amongst the expected 15 million objects to be imaged with LOFAR is also timely. 

This paper is arranged as follows. In Section~\ref{method}, the procedure for producing the training and verification data of simulated observations with the ILT is presented. Also, we present the detailed methodology of the CNN-based lens detection algorithm. Here, we test a novel deep learning component, called a convolutionalized block, which to our knowledge has not been used in lens detection applications before. In Section~\ref{tests}, we evaluate the results of the lens detection algorithm, and the lens-parameter space that the ILT is sensitive to is determined in Section~\ref{results}. Finally, in Section~\ref{discussion}, the results from this work are discussed, and concluding remarks on the methodology and future prospects are presented.

\section{Method}
\label{method}

In this section, we first summarise our pipeline for  simulating realistic gravitational lensing data from the ILT (for training and testing purposes). Then, we outline the architecture of the deep learning algorithm we use to detect and rank gravitational lensing candidates.

\subsection{Simulating a training dataset for the ILT}
\label{training_data}

\begin{figure*}
    \centering
    \includegraphics[width=14cm, height=5cm,]{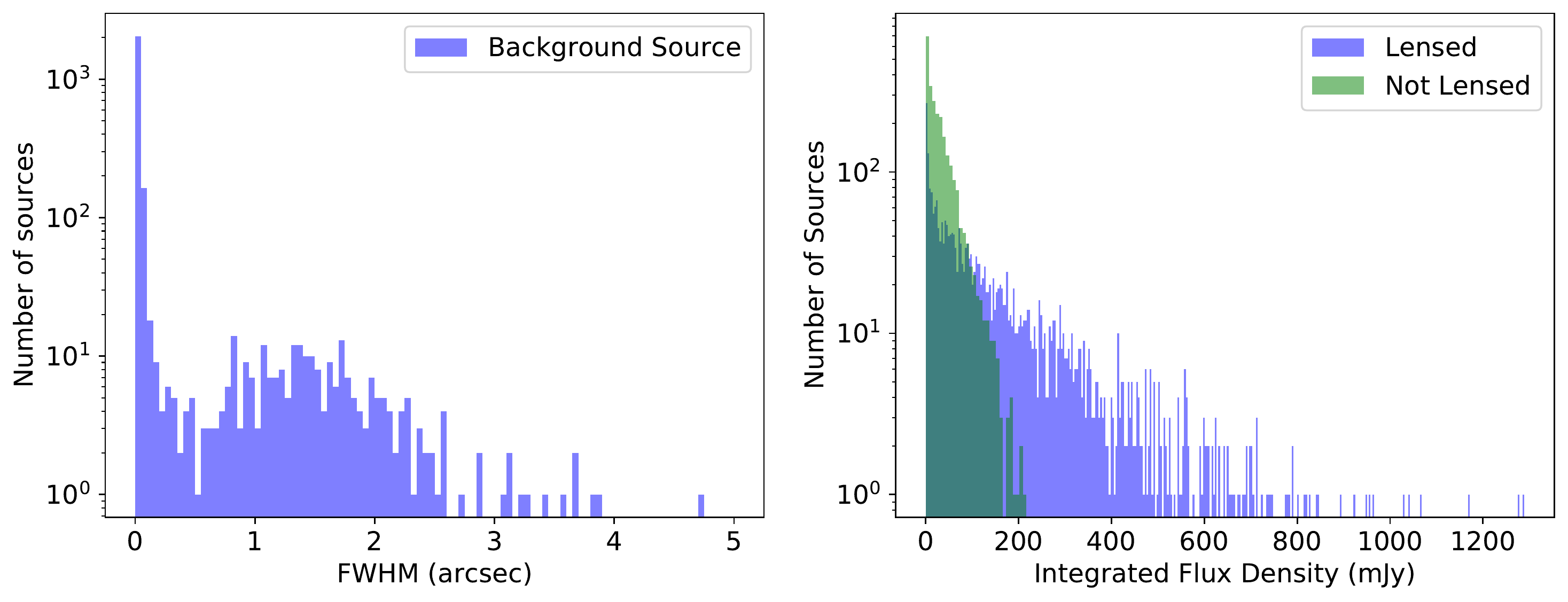}
    \caption{The left panel shows the distribution of full width at half maxima (FWHM) of the Gaussian background sources used for the training dataset. The right panel shows the integrated flux density of the background source population (green) in comparison to the integrated flux density of the lensed source population (blue). The effect of the magnification provided by the gravitational lensing is clearly seen in the source number counts as a function of flux density.}
    \label{fig:trainFHWM}
\end{figure*}

The process of simulating realistic observations of gravitational lenses with the ILT will be presented in detail by de Roo et al. (in prep.). In summary, the lensing data are created using {\sc PyAutoLens} \citep{Nightingale2021}, an open source code that generates simulated images through the process of ray-tracing. Although {\sc PyAutoLens} also has the functionality to produce interferometric datasets, here we have only incorporated the ray-tracing component in our pipeline to create a mock lens-plane model from a parametric source-plane model. We then make our own visibility datasets and produce deconvolved images using the Common Astronomy Software Applications (CASA; \citealt{McMullin2007}) package; see \citet{Rezaei2021} for a more detailed explanation of generating visibility data from mock sky-model images.

The lens simulation platform generates mock analytic sources and places them randomly on a grid. For simplicity, we have chosen to represent the background source surface brightness distribution with multiple elliptical Gaussian components. Fig. \ref{fig:trainFHWM} shows the distribution of full width at half maxima (FWHM) for the simulated background source components in our training dataset. This distribution of source sizes is based in part on the typical sizes of radio emitting star-forming galaxies at 1.5~GHz \citep{Muxlow2020}, but as most of our sources will likely be associated with AGN activity, which can have a variety of sizes, we have also added a distribution of sources with sizes $<0.5$~arcsec. This upper-limit on the AGN source size was chosen to be consistent with recent ILT observations of the Lockman Hole region, where 88 per cent of the detected sources were found to be compact at 144 MHz (beam size 0.4 arcsec; rms 25~$\mu$Jy~beam$^{-1}$; \citealt{Sweijen2022}). As gravitational lensing has only rarely been observed when the object has large-scale radio lobes \citep{Haarsma2005}, we initially chose not to densely sample source sizes larger than 2.5 arcsec (see Section \ref{fake_doubles}, where we sample non-lensed sources that are extended between 0.5 and 6 arcsec). 

The resulting distributions of the source sizes and the flux densities in our training dataset are imbalanced. As can be seen from Fig.~\ref{fig:trainFHWM}, the majority of the simulated source components in our training dataset are $< 0.5$ arcsec in size. We assume this will not affect the performance of the trained model, or the completeness and purity of the detected lens candidates. This is because we expect the proposed networks to learn the lensing properties and distinguish between lensed and non-lensed events based on the observed differences in the surface brightness distributions of both classes. We address this assumption below when analysing our final network structure. Also, as can be seen from Fig.~\ref{fig:trainFHWM}, the distribution of integrated flux density for the lensed and non-lensed classes is also imbalanced. We do not expect this to impair the quality of our analysis for two reasons. First, we apply a normalization technique to the lensed and non-lensed events in order to keep all input images in the same range of pixel values. Second, as we are interested in understanding the limitation of deep learning algorithms for lens finding applications, a wide range of signal-to-noise ratios have been used. Note that in Fig.~\ref{fig:trainFHWM}, the magnification effect of gravitational lensing can be clearly seen.

The simulated background sources could have up to three Gaussian components, drawn from the distributions given in Fig.~\ref{fig:trainFHWM}, in order to replicate a typical core and double-lobe structure, or a one-sided core-jet structure. In each case, the first component is compact and is injected at a random position from a uniform circular distribution around the lens centre, within a radius of $<0.8$~arcsec; the additional components were then added co-linearly to simulate jetted radio sources with a size of $<0.2$~arcsec. The position relative to the lens was chosen so that at least one of the radio components of the background source would likely be strongly lensed, that is, produce multiple images. Note that the maximum distance between the radio components was chosen considering the properties of radio sources observed by the ILT at 144~MHz \citep{Sweijen2022}. However, as stated above, we also tested larger separations between two radio components for the non-lensed samples to test the reliability of our approach to distinguish between double-lobed radio sources and gravitational lenses that produce two lensed images.

\begin{figure*}
\begin{center}
\includegraphics[width=\textwidth, trim=5cm 0cm 5cm 0cm]{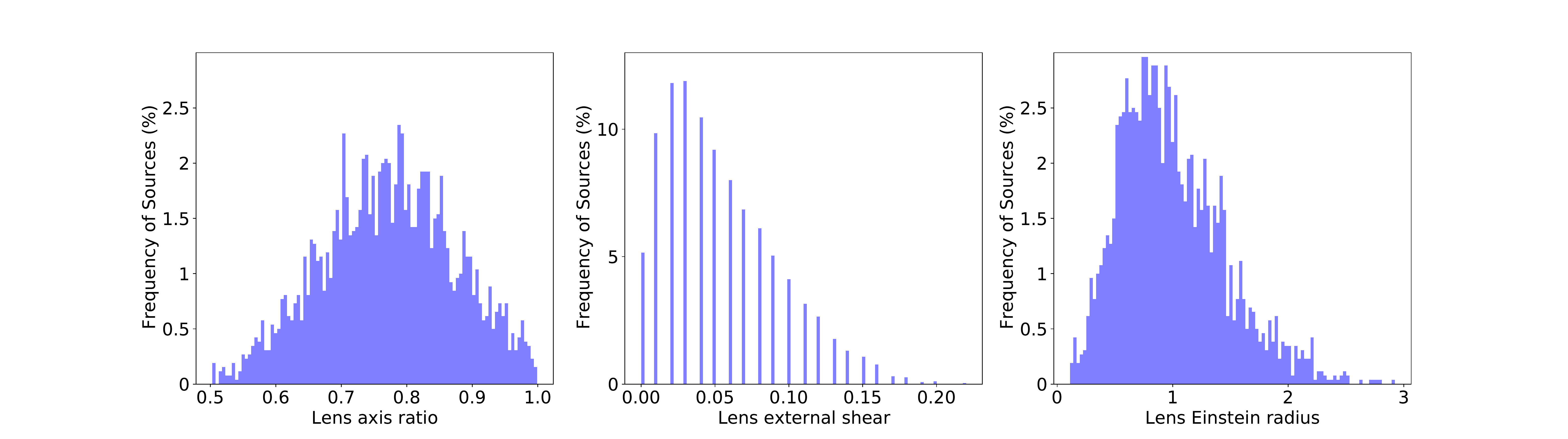}
\caption{The distribution of the lens model parameters used for the training dataset; these are (left panel) the lens axis ratio $(b/a)$, (middle panel) the lens external shear ($\gamma_{\rm ext}$), and (right panel) the lens Einstein radius ($\theta_E$). The position angles of the ellipsoidal mass distribution and the external shear were set randomly between $\pm90$~deg.}
\label{fig:lens_prop}
\end{center}
\end{figure*}

The resulting mock lensed images of the simulated sources were then generated using a singular isothermal ellipsoid (SIE) mass model, with an external shear contribution. Such a model has been shown to be a good approximation for the mass distributions of massive elliptical galaxies (e.g. \citealt{Koopmans2006}). The parameters used to describe this mass model are the lens position (always at the centre of the grid), the axis ratio ($b/a$) and position angle of the ellipsoid, the shear strength ($\gamma_{\rm ext}$) and position angle, and the Einstein radius ($\theta_{E}$), which is used as a proxy for the lensing mass. For our initial tests (see Section~\ref{tests}), the mass models were generated using a combination of these parameters drawn from the distributions of real gravitational lens mass models \citep{Bolton2008}, except for the Einstein radius where the image-separation distribution for all known lensed quasars was used (the Einstein radius is approximately half of the maximum image separation). The resulting distributions for these lens model parameters are shown in Fig.~\ref{fig:lens_prop}. However, for our final tests (see Section \ref{results}), we use a uniform distribution for the Einstein radius and the axis-ratio to obtain an unbiased assessment of the network's ability to identify gravitational lenses. Throughout, we only search for lens systems with Einstein radii $\geq 0.15$~arcsec, to  exclude part of the model space that the ILT is likely not sensitive to, given the resolution of the data. For simplicity, we fixed the lens redshift to be $z_l = 0.5$ and the source redshift to be $z_s = 1.0$. This choice will have no impact in our simulations since it is the angular-scale, as set by the varying Einstein radius that matters.

In the next step, we generated realistic interferometric visibility datasets, which was done by using CASA. First, we used an actual observation with the ILT of the bright radio galaxy 3C\,330 (Lafontaine et al. in prep.), to provide a skeleton MeasurementSet with the correct number of stations, {\it uv}-coverage and visibility averaging time and bandwidth. This simulation consisted of the phased-core station, fourteen remote stations, and thirteen international stations of LOFAR. In our simulations, we assume that we have perfectly calibrated data or that the calibration errors are relatively small compared to the thermal noise. We will address the effect of calibration errors on the lens properties in a future study.

The total time on-source was set to 8 h, which is typical for an observation that is taken as part of the LOFAR Two Metre Sky Survey (LoTSS; \citealt{Shimwell2019}). Each lensed and non-lensed model surface brightness distribution was sampled in the visibility plane via a Fourier transform, and the resulting visibilities were then corrupted by adding Gaussian noise such that the final images had an rms noise of 90~$\mu$Jy~beam$^{-1}$ (Briggs weighting; ${\rm Robust} = 0.5$; {\it uv}-range $> 80$k$\lambda$); this is the typical rms noise for an ILT observation \citep{Morabito2021}. The resulting beam size was $379\times293$~mas at a position angle of $-16.9$ deg East of North. The 
%5200 
simulated visibility datasets were then imaged and deconvolved using the {\sc tclean} task within CASA. The final images have $64\times64$~pixels, where each pixel is 0.12 arcsec in size. This is equivalent to a sky-area of $7.68\times7.68$~arcsec$^2$ or $4.55\times10^{-6}$~deg$^2$. With our current understanding of the density of radio sources observed with ILT at 144 MHz to the depth of a typical LoTSS observation, the probability of having two unrelated radio sources being located within the sky-area of the simulated fields is about $1.5\times10^{-3}$. Therefore, we do not consider the presence of unrelated radio sources within the fields-of-view to be an issue (although we do consider non-lensed radio sources with multiple components in our analysis, see below).

Fig.~\ref{fig:sample_strong_lenses} shows a subset of the simulated gravitational lens systems and non-lensed radio sources as if they were observed with the ILT as part of LoTSS. We note that our final training dataset includes different types of gravitational lens systems, such as those with extended emission, like gravitational arcs and Einstein rings, and those with compact lensed emission producing two or four images of the background source. We have also included a wide range of signal-to-noise ratios to evaluate the performance of our lens detection algorithm when the source surface brightness is close to the rms noise level of the data.

\begin{figure*}
\begin{center}
\begin{minipage}[b]{\textwidth}
\centering
\includegraphics[height=2.35cm]{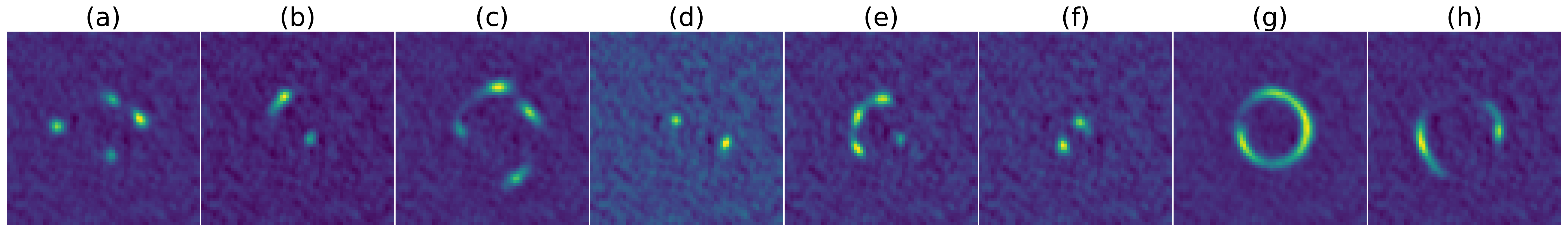}
\end{minipage}
\begin{minipage}[b]{\textwidth}
\centering
\includegraphics[height=2.35cm]{Fig3_2.pdf}
\end{minipage}
\newpage
\caption{A selection of strong gravitational lens systems (upper row) and non-lensed radio sources (lower panel), as if observed with the ILT as part of LoTSS. Each image contains $64\times64$~pixels and is equivalent to a sky-area of $7.68\times7.68$~arcsec$^2$. The synthesized beam size is $379\times293$~mas at a position angle of $-16.9$~deg East of North, and the rms noise is 90~$\mu$Jy~beam$^{-1}$ (natural weighting).}
\label{fig:sample_strong_lenses}
\end{center}
\end{figure*}

\subsection{Network structure}

\begin{figure}
    \centering
    \includegraphics[width=11cm, trim=3cm 3cm 3cm 2cm]{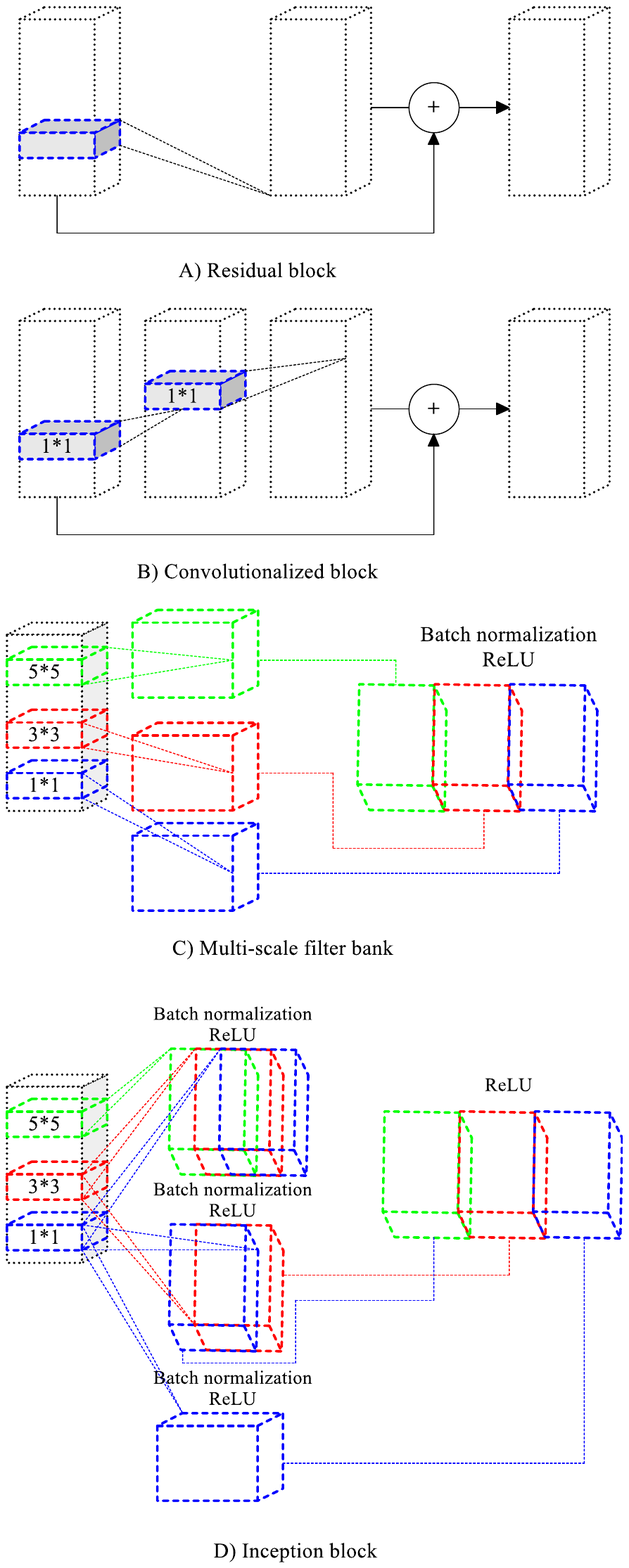}
    \caption{A block diagram showing the main components that have been used to construct the three CNNs tested here to detect and rank  gravitational lens candidates in ILT imaging data.}
    \label{fig:myblocks}
\end{figure}

In this section, we present our CNN based approach to detect and rank strong gravitational lens systems from ILT-quality imaging data at 144 MHz. CNNs use several layers to process the input imaging data of a topological structure, and they have become the prominent approach for object detection and classification with machine learning. Each layer has a unique functionality, and different arrangements of the layers can make a variety of convolutional components. Therefore, the performance of a CNN is dependent on the implemented components, and each component may perform differently for a specific application. We have implemented four main components;  a residual block, a convolutionalized block, a multi-scale filter bank and an inception block. These components are being employed in three separate network structures, which we call structure 1, 2 and 3. The three structures are then trained with the same training dataset and their performance is evaluated using the same test and validation datasets. In the following, we first provide our motivation for using each of the four main components, before outlining the arrangement of these components to build the three network structures.

The training of the networks is guided by the gradient based optimization of a suitable loss function. Here, the gradient refers to the change of the loss function with respect to the weights in the network. For networks with many convolutional layers, the magnitude of the gradient in earlier layers can become small due to the multiplications imposed by the chain rule. As a result, updates are diminutive and the progress of the training can be slow. This "vanishing gradient problem" can occur faster when the number of layers in the network is increased, and it motivated the idea of using "residual blocks" to limit its effect \citep{He2015}. Part A of Fig.~\ref{fig:myblocks} shows the concept of residual blocks. 

A convolutionalized block is presented in part B of Fig. \ref{fig:myblocks}. It has $f$ convolutional filters, each with a filter size of ($1\times 1 \times d$), where $d$ is the depth of the input image. As we currently use only single images produced from interferometric datasets, there is only one dimension containing information on the source surface brightness, and therefore, $d$ is equal to 1. In the future, multiple images that take into account the source surface brightness distribution as a function of frequency will be implemented; in this case $d>1$. A convolutionalized block with $f$ filters is considered to be equivalent to fully connecting the input object of $1 \times 1 \times d$ to $f$ output nodes \citep{Lee2017}. Residual learning is implemented in the convolutionalized blocks to improve the training efficiency when extracting source components in feature space. Besides the  convolutionalized block, we have also adapted the concept of a multi-scale convolutional filter bank (part C in Fig.~\ref{fig:myblocks}) from the same study \citep{Lee2017}. It is used to simultaneously scan through local regions of the input image and then exploit various local spatial structures. It then concatenates the extracted feature maps to be used together. A similar concept is employed in an inception-based block, which was first introduced in GoogLeNet by \citet{Szegedy2016}, and provides a deep, but also a wide structure that allows several independent paths in the model to be optimized. The block in part D of Fig.~\ref{fig:myblocks} is inspired from the inception block and provides three paths from the input. The first path starts with a filter of size ($1\times1$), while the second path applies another convolution with a filter of size ($3\times3$), after convolving with the filter of size ($1\times1$). The last path contains both ($1\times1$) and ($3\times3$) sized filters that are followed by a ($5\times5$) sized filter bank. This technique has been implemented due to the varying size and structure of the gravitational lens systems that we are interested in detecting. Due to the dependence of the morphology of the lensed images to the different input lens and source parameters, it is important to build a flexible model that can handle different configurations of the lensed images.

\begin{figure*}
\begin{center}
\includegraphics[height=6.5cm]{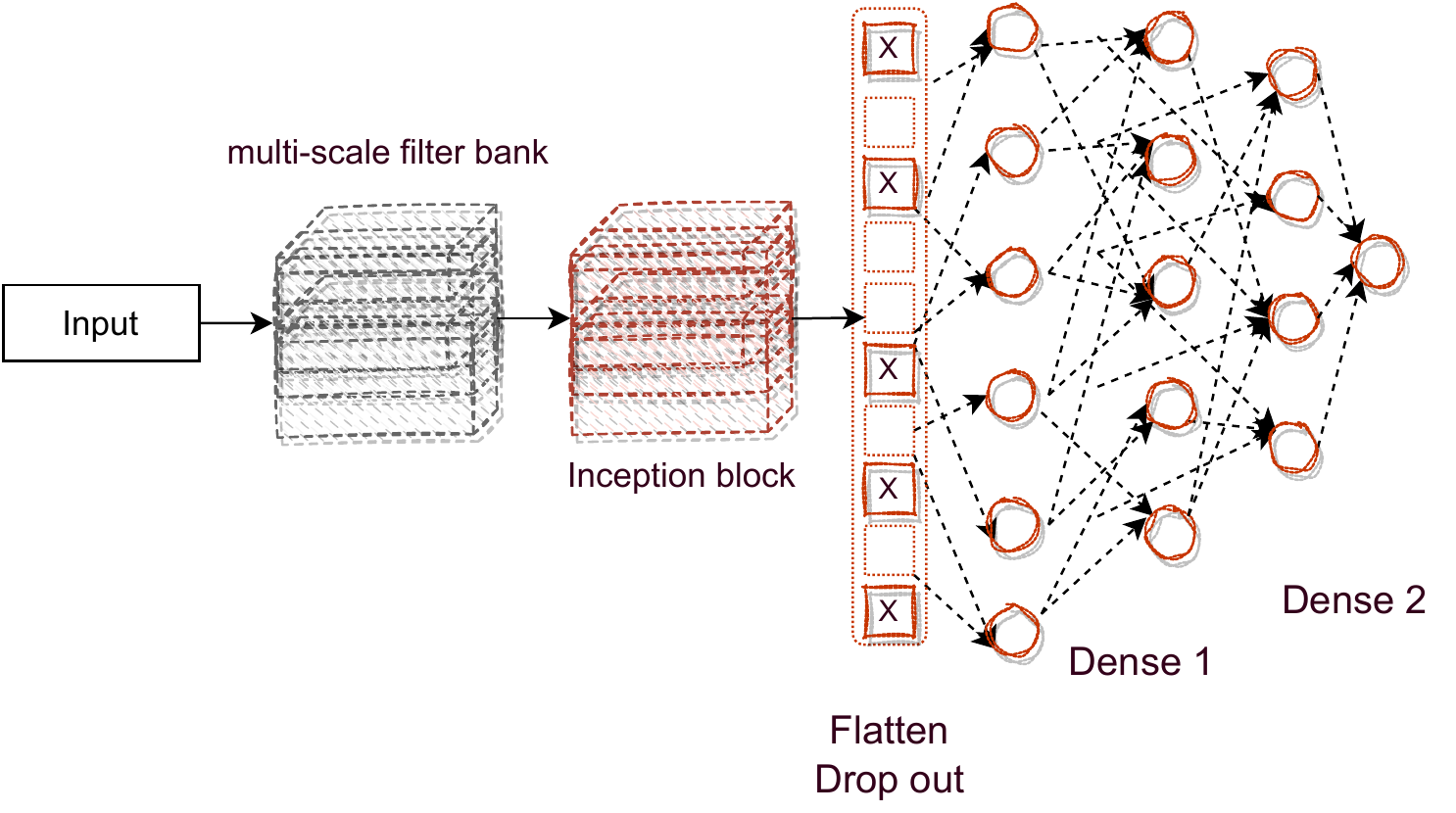}
\caption{Structure 1: This network structure consists of a multi-scale filter bank and inception blocks that are placed next to each other. The output of the inception block is then sent to the dropout layer, which is connected to two dense layers of fully connected networks. There are in total 6\,480\,753 trainable parameters in this architecture.}
\label{fig:arc1}
\end{center}
\end{figure*}

\begin{figure*}
\begin{center}
\includegraphics[height=6.3cm]{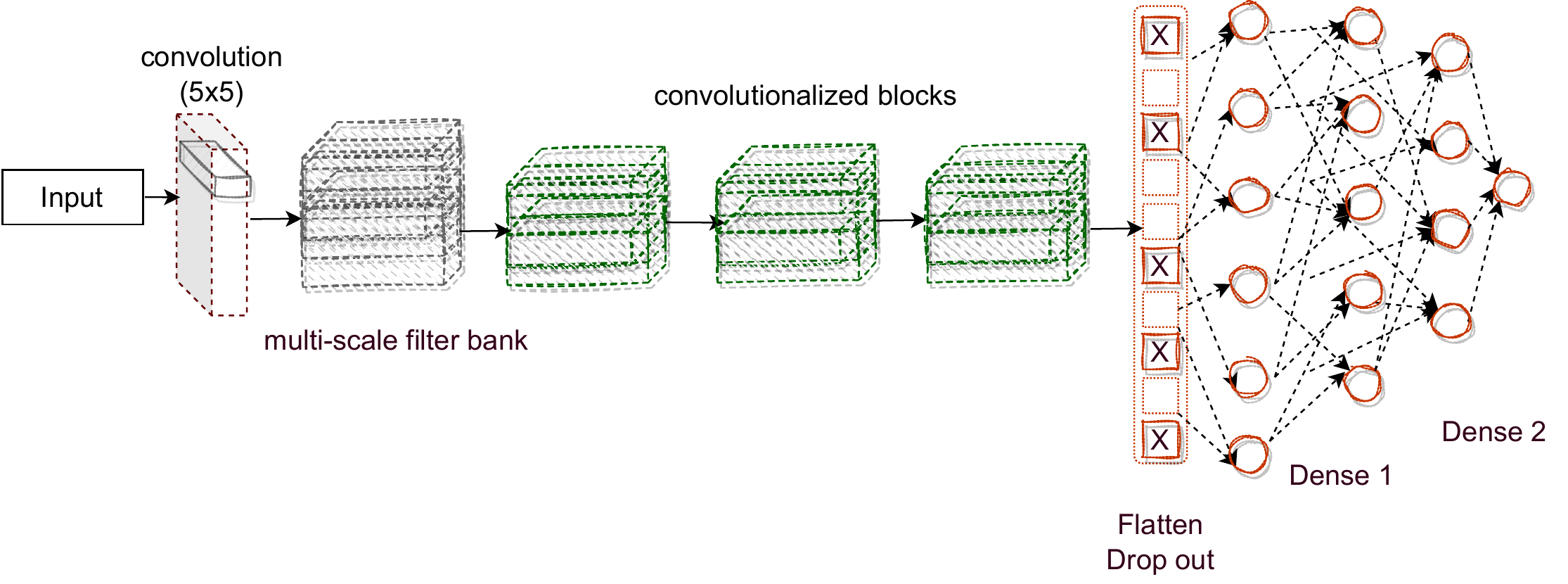}
\caption{Structure 2: For this structure the input is convolved with 128 filters of size ($5\times5$) before a multi-scale filter bank is applied. Next, three convolutionalized blocks are placed, where the residual blocks are used with three convolutional layers. Similar to structure 1, a dropout layer and two fully connected layers are implemented to extract the output of the network. The total number of trainable parameters in this structure is 4\,538\,497.}
\label{fig:arc2}
\end{center}
\end{figure*}

\begin{figure*}
\begin{center}
\includegraphics[height=5.4cm]{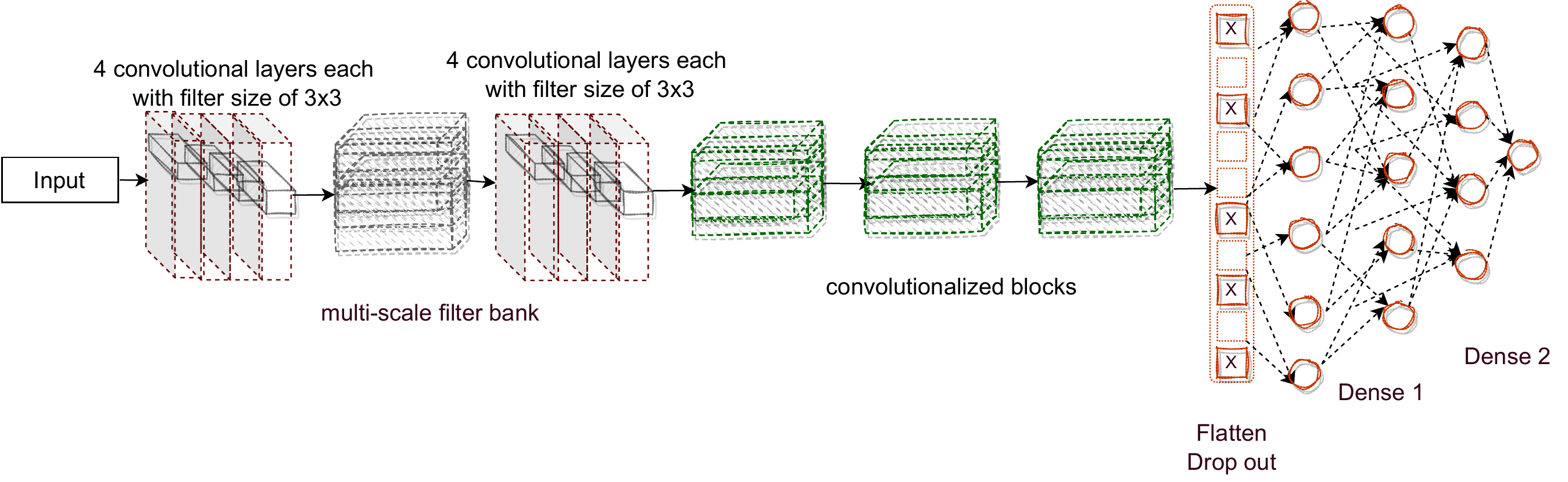}
\caption{Structure 3: In addition to the convolutionalized blocks and multi-scale filter bank used for structures 1 and 2, this structure also consists of two sets of 4 sequential convolutional layers, each with a filter size of ($3\times3$). With 7\,273\,813 trainable parameters, this structure has the largest number of parameters.}
\label{fig:arc3}
\end{center}
\end{figure*}

Considering the main components shown in Fig.~\ref{fig:myblocks} as the building blocks, we now present the three different network structures that are trained to separate and rank images of gravitational lenses from those of non-lensed samples. Fig.~\ref{fig:arc1} shows the first network, structure 1, which contains a multi-scale filter block followed by an inception block. A dropout layer \citep{Gal2015} followed by two fully connected layers is placed at the end. The dropout layer randomly removes units with a probability of $p_d$ at each step during training, without updating the weights of those units. This helps to prevent overfitting by reducing the effective network complexity. The second network,  structure 2, is displayed in Fig.~\ref{fig:arc2}. It contains a convolutional layer with a filter size of ($5\times5$), followed by a multi-scale filter bank and a set of three convolutionalized blocks. At the end, there are two fully connected layers with a dropout function. Finally, in Fig.~\ref{fig:arc3} we present the third implemented network, structure 3, which uses two sets of four convolutional layers, each with a filter size of ($3\times3$). The two sets of convolutional layers are connected via a multi-scale filter bank. Similar to structure 2, this network contains three convolutionalized blocks, followed by two fully dense layers and a dropout layer. 

\subsection{Preprocessing} \label{preprocessing}

Preprocessing of the imaging data is important for optimizing the performance of the learning algorithms. Here, we keep the range of pixel values for both the lensed and non-lensed samples in the same range using a MinMax normalization, such that
\begin{equation}
\centering
    x_{\rm normalized}= \frac{x-\min(x_d)}{\max(x_d)-\min(x_d)},
\label{eq:minmax}
\end{equation}
and where all pixel values in a given image are mapped to the range $(0,1)$. Here, $x$ is the value of a given pixel and $x_d$ represents all the pixels in the image. 
%Input normalization helps to regularize the model by harmonizing the distribution of inputs. 
Note that, due to surface brightness correlations introduced by the lens equation, it is the relative surface brightness of the lensed images (and non-lensed source emission) within each simulated sample that matters. Therefore, losing the absolute amplitude scaling of the data with this form of normalization will have no affect on our ability to identify lens candidates.
 
\subsection{Loss function}

Loss functions are designed to measure the performance of a network in terms of the dissimilarity between the estimated and true class labels. This difference is then optimized using the gradient based Adam optimization algorithm \citep{Kingma2014}. Since we perform a binary classification between lensed and non-lensed objects, we have chosen the Binary Cross Entropy (BCE) form of loss function for our tests. It is given as,
\begin{equation}
\centering
\begin{split}
{\rm BCE} = -\frac{1}{N} \sum_{i=1}^{N} y_i\log p_i + (1-y_i)\log (1-p_i).
\end{split}
\label{eq:bce}
\end{equation}
For the BCE loss function, we consider a given class label $y_i$ 
%and a predicted label $y'$ 
in a dataset of $N$ training samples in each batch $i$, where $p_{i}$ is the estimated probability of the input image being a lens and $1-p_{i}$ is the probability of it not being a lens.

\subsection{Detection probability uncertainty} \label{model_uncertainty}

A key goal of our lens identification algorithm is to also give some confidence on the level of detection. This is important because even though we may expect to identify many candidates with the ILT (and the ngVLA and SKA in the future), confirming the gravitational lensing hypothesis will also require additional telescope time. Therefore, it is important to have a ranked list of candidates for prioritising any follow-up observations, with some estimate of the uncertainty on the lensing probability. This is why we have also implemented a dropout component \citep{Gal2015} in our three network structures (see Figs.~\ref{fig:arc1} to \ref{fig:arc3}). This method imitates Bayesian models within the deep learning algorithm without changing the network structure and the basic method of optimization.

Traditionally, dropout has been applied in deep learning algorithms to prevent over-fitting and to regularize the model (e.g. \citealt{Srivastava2014}). During the training phase, dropout is used by randomly removing units and their corresponding connections from the network with some probability $p_d$. By applying dropout, a new set of neurons is eliminated at each iteration, resulting in a virtual change to the network's structure. This provides a way of approximately combining many different neural networks together with shared weights. The ratio of eliminated units at each iteration is defined by $p_d$. For a wide range of networks and applications, $p_d=0.5$ tends to generate close to optimal results \citep{Srivastava2014}. 

We have also implemented a Monte Carlo dropout approach, which can be described as the approximate integration of dropout over the weights of the models. It is based on running the network for some number of iterations on a test dataset and building a distribution of the predicted probabilities. The mean and standard deviation of this distribution contain information about the most reliable prediction, as well as the model uncertainty. Dropout has become a popular method to measure model uncertainties in gravitational lensing applications of deep learning \citep{Hezaveh2017,Perreault-Levasseur2017,Bom2019,Maresca2021}. 

\subsection{Evaluation criteria}

Defining proper classification criteria is needed to interpret and compare the results of the three different network structures tested here. Since the lens detection problem can be defined as a binary classification (i.e. lensed or non-lensed), we use the reference confusion matrix presented in Table~\ref{tab:sample_confusion} to evaluate our results. Here, a true positive (TP) is defined as a genuine gravitational lens system that has been successfully classified as such, while a true negative (TN) is defined as when a non-lensed source has been correctly identified as not being a gravitational lens. Conversely, a false positive (FP) occurs when a non-lensed source is categorised as a gravitational lens, and a false negative (FN) corresponds to those gravitational lensing events that are missed by the algorithm, and are classified as non-lensed sources. An ideal network would detect all gravitational lens systems and reject all non-lensed samples. However, this can hardly ever be achieved in a practical real world problem due to noise in the data.
 
We evaluate the performance of the three network structures by using the following criteria: accuracy, precision, recall and fall~out. 

Accuracy is the most trivial metric to determine as it is the total number of correct predictions for both the lensed and non-lensed classes divided by the size of the test dataset. Precision measures the ratio between the correctly identified gravitational lenses and all samples identified as gravitational lenses. Recall on the other hand measures the ratio of the detected gravitational lenses to the number of lensing events in the test dataset. This is also known as the completeness or TP rate. Fall-out, or the FP rate, measures the ratio of misidentified gravitational lenses to the total number of non-lensed events in the test dataset. They are each calculated using,  

\begin{equation}
\centering
\begin{split}
{\rm Accuracy}~= \frac{\rm TP+TN}{\rm TP+FN+TN+FP},
\end{split}
\end{equation}
\begin{equation}
\centering
\begin{split}
{\rm Precision}~= \frac{\rm TP}{\rm TP+FP},\quad 
%{\rm precision_{Not-lens}}~= \frac{\rm TN}{\rm TN+FN},
\end{split}
\end{equation}
\begin{equation}
\centering
\begin{split}
{\rm Recall}~= \frac{\rm TP}{\rm TP+FN}, \quad
%{\rm recall_{Not-lens}}~= \frac{\rm TN}{\rm TN+FP}.
\end{split}
\end{equation}
\begin{equation}
\centering
\begin{split}
{\rm Fall~out}~= \frac{\rm FP}{\rm FP+TN}.
%{\rm recall_{Not-lens}}~= \frac{\rm TN}{\rm TN+FP}.
\end{split}
\end{equation}
Typically, the success of a gravitational lens search algorithm is not judged purely on recall, as given the large number of gravitational lenses that are expected to be found, completeness is not important for most science goals. Instead, the quality tends to be judged on a low fall-out, as we ideally want to have a high level of genuine lens candidates in our ranked list.

Another evaluation metric is the Receiver Operating Characteristic (ROC) curve. It is used when the performance of a binary classifier can be measured as a function of some cut-off threshold. A ROC curve shows the TP rate (recall) as a function of the FP rate (fall~out). In our lens detection algorithm, we have defined a threshold based on the output of the network, which is used to separate lensed and non-lensed samples. The choice of this threshold has a significant impact on all the evaluation metrics listed above, which we discuss in the next section. The ROC curve provides a qualitative measure that is independent of a specific threshold, and is therefore, also a useful metric to consider.

\begin{table}
    \centering
\begin{tabular}{l|l|c|c|}
\multicolumn{2}{c}{}&\multicolumn{2}{c}{True Data}\\
\cline{3-4}
\multicolumn{2}{c|}{}&Lens&Not Lens\\
\cline{2-4}
{Test Results}& Lens & TP & FP\\
\cline{2-4}
& Not Lens &FN & TN \\
\cline{2-4}
\multicolumn{1}{c}{} & \multicolumn{1}{c}{Total} & \multicolumn{1}{c}{Lens Sources} & \multicolumn{    1}{c}{Normal Sources} \\
\end{tabular}
    \caption{The sample confusion matrix, showing how the TP, FP, FN and TN are defined.}
    \label{tab:sample_confusion}
\end{table}

\section{Network tests}
\label{tests}

In this section, we present the results of applying the three network structures to a set of test data, which were created in a similar way to the training dataset described in Section~\ref{training_data}. First, we present the criteria used to determine whether a sample has been identified as a lensed or non-lensed event. Next, we investigate three scenarios that have been designed to evaluate the performance of each network structure. The first scenario uses a test dataset with a realistic distribution of lens model parameters, which allows us to test the network performance with a dataset that is similar to the training dataset.
The second scenario contains test samples generated with a uniform population of lens model parameters. This allows us to test the performance of each structure to exotic lens configurations that were not necessarily well-represented in the training data. For these two tests, we trained the networks using 3000 model lensed images, to represent the class of lensed objects, and 3000 corresponding source models to represent the class of non-lensed objects (generated as described in Section~\ref{training_data}). In total, a maximum of 450 training epochs were used. 

Our third test is designed to investigate the ability of the three networks to correctly label non-lensed double-lobed radio sources. For this test, we augmented the training data set with 2000 simulated double-lobed radio sources. The test datasets for each experiment are described below.

\subsection{Determining the lens probability}

A lensing probability in the range (0, 1) is predicted by the output of the three network structures, where the closer the output is to 1, the higher the probability of the given sample being a gravitational lens. Since we have used dropout in the training, we are able to calculate the model confidence for a specific prediction (see Section 2.5). When dropout is also active during the testing mode, the network randomly eliminates some of the connections and makes the best predictions based on the available data. Not having access to the entire trained set of weights and neurons leads to varying outputs for a given sample during testing. To calculate the most reliable output, we averaged the network prediction over 250 realizations for each single test sample to calculate the final probability. This approach also provides additional information by calculating the standard deviation of the probability.

Considering the output of the network as a continuous number between 0 and 1, we need to define a threshold according to which test results are classified into lensed and non-lensed samples. The value of the threshold affects all evaluation criteria such as completeness and number of false detections. In general, higher threshold values yield a lower detection of true lens samples, as the criteria are stricter, but this can also lower the number of false detections. We determine this threshold from inspecting the ROC plot for each network structure.

\subsection{Test on a realistic lensing population} \label{realistic_section}

\begin{figure}
    \centering
    \includegraphics[width=\columnwidth]{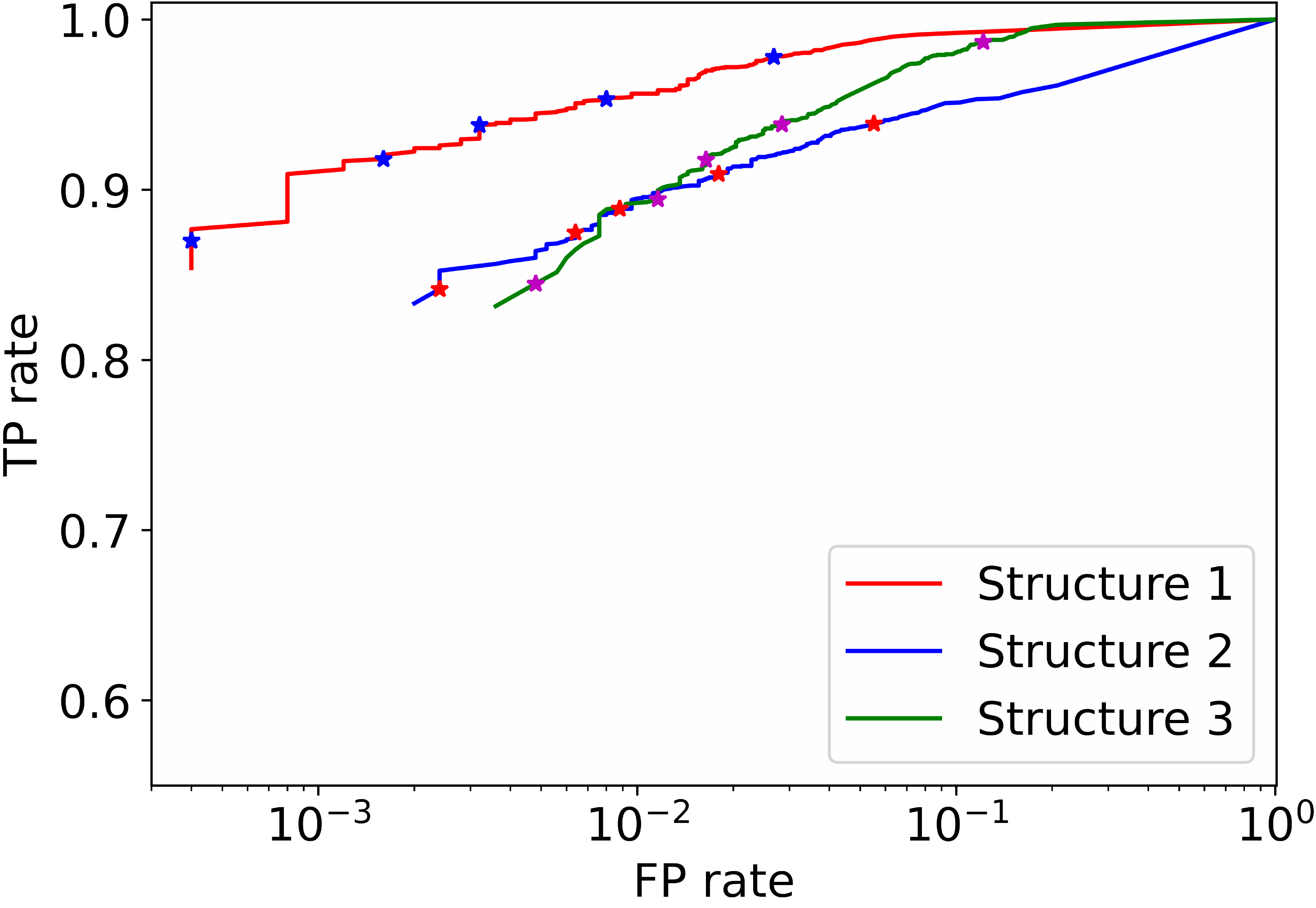}
    \caption{The TP rate as a function of FP rate for the three network structures, when applied to a test dataset created using a realistic distribution of lens model parameters. The stars correspond to thresholds of 0.1, 0.5, 0.75, 0.9 and 0.99, from right to left.}
    \label{fig:FP_TP_realistic}
\end{figure}

In the first testing mode, we have generated a realistic dataset with the same lens model parameters as our training data, shown in Fig. \ref{fig:lens_prop}. 
Here, we test a dataset containing 5000 samples, where 2500 samples are non-lensed events and 2500 samples are lensed events that are created from a realistic distribution of lens models. The results for this test are shown in the ROC plots presented in Fig. \ref{fig:FP_TP_realistic}. Overall, network structure 1 outperforms the two other structures in both the TP- and FP-rates at all probability thresholds. It has a TP rate of 95.3 per cent for a threshold of 0.5, while the FP rate (0.8~per cent) is significantly lower compared to structures 2 and 3. Although the TP rates of structures 2 and 3 are high ($>90$~per cent) for a threshold of 0.5, they have a higher FP rate (1.8 and 2.8 per cent, respectively).

Table~\ref{tab:evaluation_realistic} contains quantitative information on the defined evaluation metrics. These metrics are calculated by considering a threshold value of 0.99, which results in lowering the TP rates by between 6.8 and 9.4~per cent, while lowering the FP rates by around an order of magnitude, when compared to the results for a threshold of 0.5. We find that network structure 1 has the best performance, with TP and FP rates of 87 and 0.04~per cent, respectively, when a realistic distribution is used for the lens model parameters and a threshold of 0.99 is applied.

\begin{table}
    \centering
\begin{tabular}{
%>{\columncolor[gray]{0.9}}
llll}
&Structure 1&Structure 2 & Structure 3\\
\cline{1-4}
Accuracy& 0.957 & 0.934 & 0.943\\
{Precision}& 0.9995 & 0.9971 & 0.9943\\
{Recall}&0.870& 0.842 & 0.845\\
{Fall out}& 0.0004 & 0.0024 & 0.0048\\
\cline{1-4}
\end{tabular}
    \caption{The evaluation criteria of the three network structures when a threshold of 0.99 is used, and a test dataset created using a realistic distribution of lens model parameters is used.}
    \label{tab:evaluation_realistic}
\end{table}

\subsection{Test on a lens population of uniformly selected model parameters } \label{uniform_section}

\begin{figure}
\begin{center}
\includegraphics[width=\columnwidth]{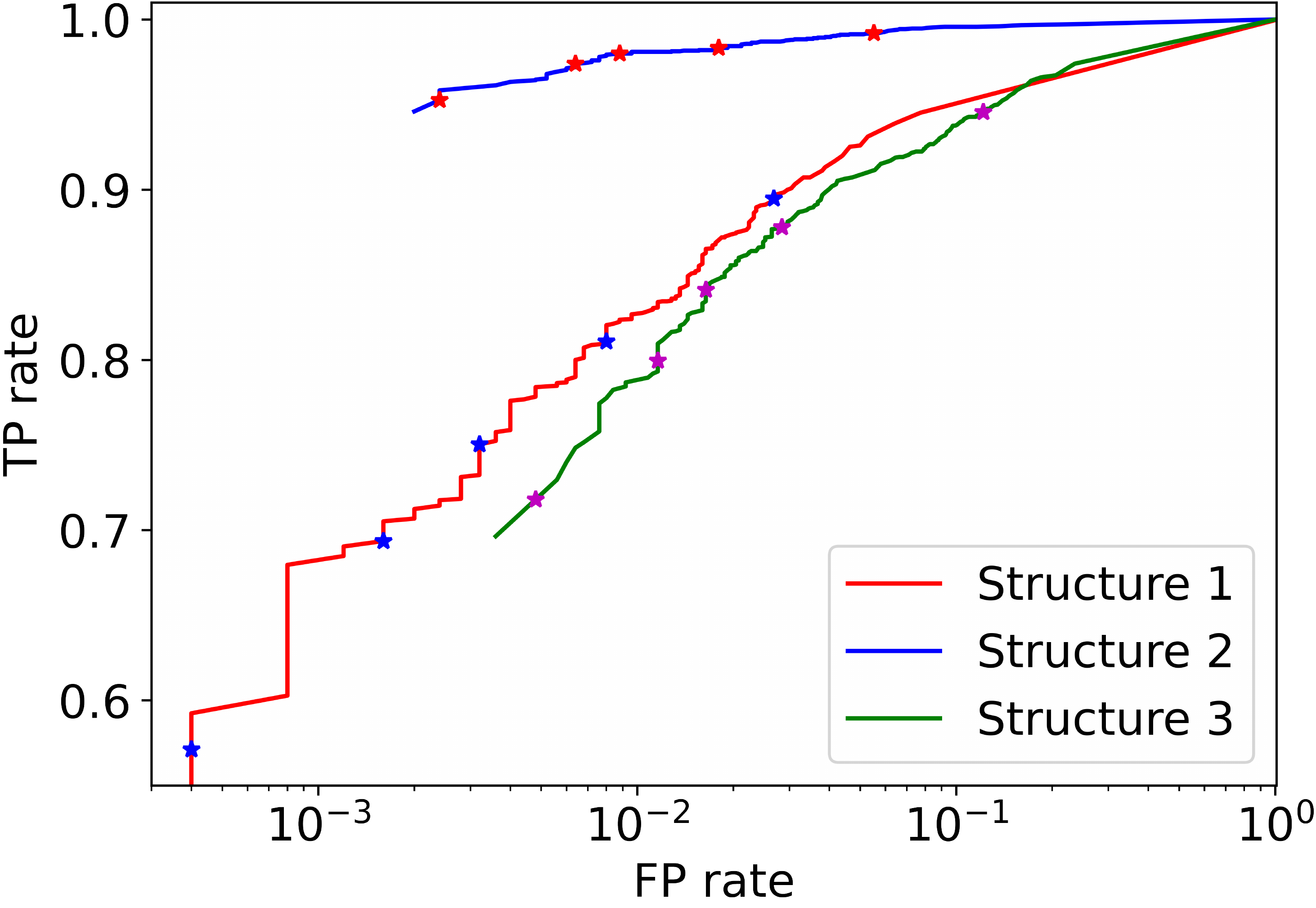}
\caption{The TP rate as a function of FP rate for the three network structures, when applied to a test dataset created using a uniform distribution of lens model parameters. The stars correspond to thresholds of 0.1, 0.5, 0.75, 0.9 and 0.99, from right to left.}
\label{fig:FP_TP_uniform}
\end{center}
\end{figure}

\begin{table}
    \centering
\begin{tabular}{
llll}
&Structure 1&Structure 2 & Structure 3\\
\cline{1-4}
Accuracy& 0.785 & 0.975 & 0.856\\
{Precision}& 0.9990 & 0.9974 & 0.9933\\
{Recall}&0.571& 0.952 & 0.718\\
{Fall out}& 0.0004 & 0.0024 & 0.0048\\
\cline{1-4}
\end{tabular}
    \caption{The evaluation criteria of the three network structures when a threshold of 0.99 is used, and a test dataset created using a uniform distribution of lens model parameters is used.}
    \label{tab:evaluation_uniform}
\end{table}

\begin{figure*}
\begin{center}
\begin{minipage}[b]{\textwidth}
\centering
\includegraphics[height=2.35cm]{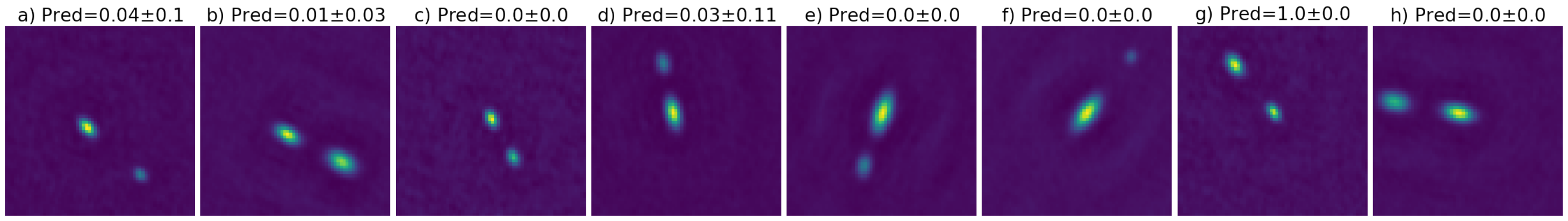}
\end{minipage}
\begin{minipage}[b]{\textwidth}
\centering
\includegraphics[height=2.35cm]{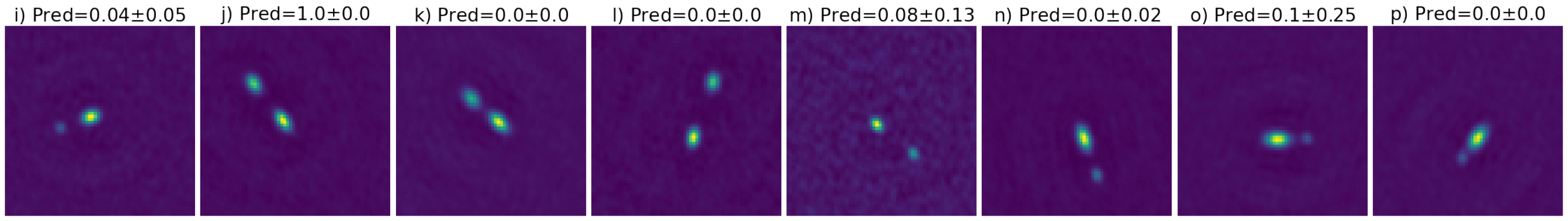}
\end{minipage}
\newpage
\caption{A representative sample of simulated (non-lensed) double-lobed radio sources. The second component is injected randomly within 3 arcsec distance of the first component. The size of the two components are drawn randomly from the distribution shown in Fig.~\ref{fig:trainFHWM}. The average predicted probability and its uncertainty over 250 iterations for structure 2 is stated above each image. Each image contains $64\times64$~pixels and is equivalent to a sky-area of $7.68\times7.68$~arcsec$^2$.}
\label{fig:fakedoubles}
\end{center}
\end{figure*}

We now provide the test results when the trained models are applied to a test dataset using a uniform distribution of lens model parameters. Similar to the previous case, the test dataset contains 5000 samples, of which 2500 samples are lensed events and 2500 are non-lensed events. Note that the set of non-lensed radio sources is the same as used above, therefore, the FP rates are unchanged. The generated test dataset has a uniform distribution for the Einstein radius (between 0.15 and 2.8 arcsec) and the axis ratio (between 0.05 and 1). This was done to ensure that each bin in the parameter space was equally sampled, which makes the sensitivity of the three networks, as a function of the lens model used, clearer to judge. Note that we draw the shear strength from the realistic distribution, as extremely high values of the shear lead to unusual and highly exotic lensed image configurations (particularly when the axis ratio of the lens is low). As with the training dataset, we also set the position angle of the ellipsoid mass distribution and the shear to be uniformly sampled between $\pm90$~deg. 

The ROC plots of the three network structures, with the probability thresholds labelled, are shown in Fig.~\ref{fig:FP_TP_uniform}. Comparing the results with those shown in Fig.~\ref{fig:FP_TP_realistic}, we find a significant drop in the TP rate at all thresholds for network structures 1 and 3, while network structure 2 performs better. The reason for this is likely due in part to each network being able to label lensed features in a different way. For example, structure 2 is better at identifying the type of lens systems generated by a uniform distribution, that is, those with smaller Einstein radii or axis ratios (see below). 
In Table \ref{tab:evaluation_uniform} we present the evaluation criteria when the trained models are tested on a dataset with a uniform distribution for the lens model parameters, and a threshold of 0.99 is applied. Compared to the previous results, we see that the TP rate of network structure 1 has decreased by 29.9 per cent, and for network structure 2 has decreased by 12.7 per cent. These results highlight that network structures 1 and 3 require the training data set to be representative of the analysis data. For network structure 2, the TP rate has increased by 12 per cent, when compared to the previous case, and now has a value of 95.2 per cent.

\subsection{Non-lensed double-lobed radio sources}
\label{fake_doubles}

\begin{figure}
    \centering
    \includegraphics[height=5cm]{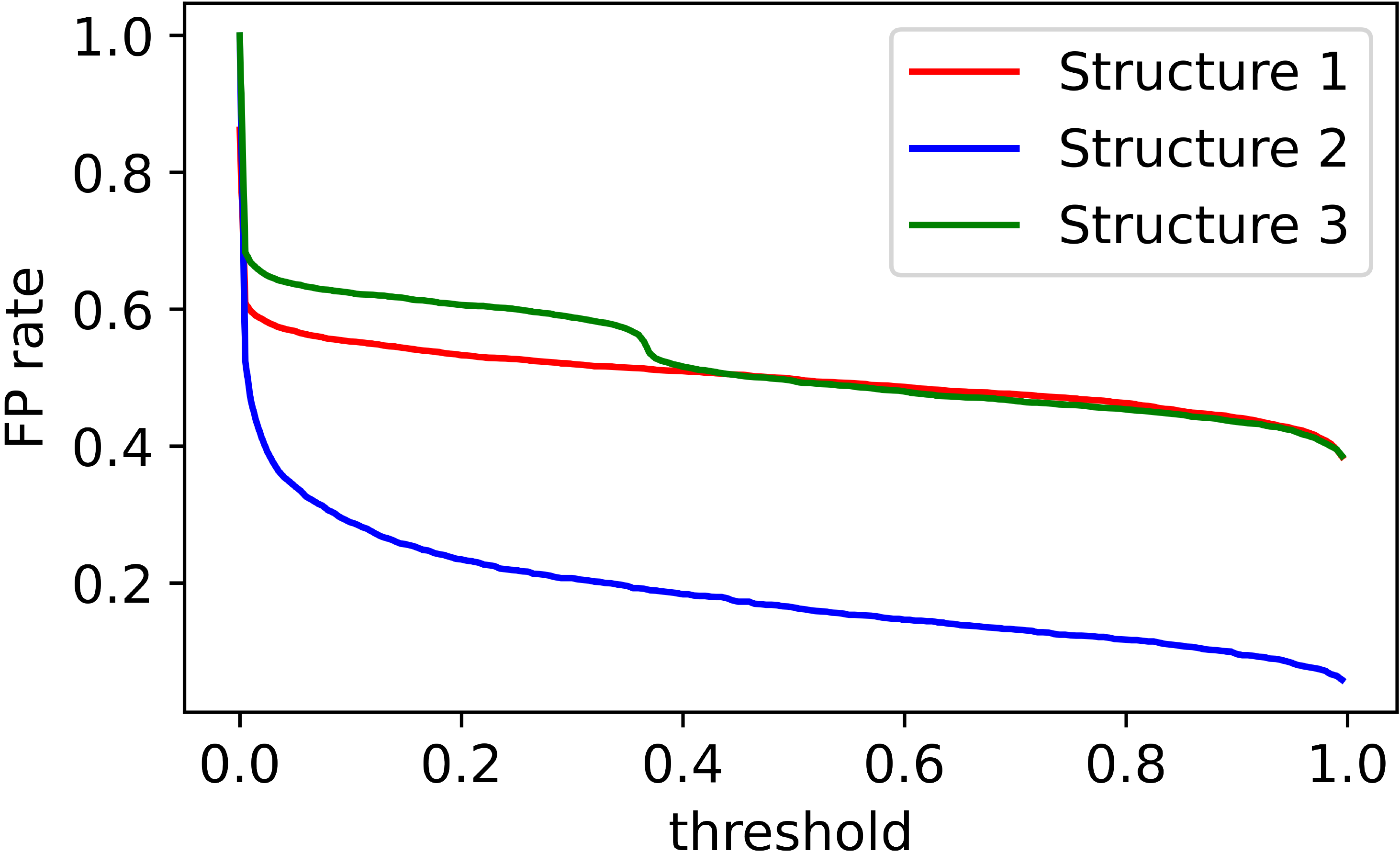}
    \caption{The FP rate as a function of threshold for a test dataset that only includes non-lensed double-lobed radio sources. Note that these types of sources were not included within the training dataset.}
    \label{fig:FP_fakedoubels}
\end{figure}

As discussed above, one of the main goals of a lens detection algorithm is to limit the number of FP events. The results for when both a uniform and a realistic distribution for the lens model parameters (Einstein radius and axis ratio) is used yields a low FP rate of 0.04 per cent for network structure 1 (equivalent to 1 FP event in our test dataset of 2500 non-lensed sources). This is due to the lensed emission having a very distinctive arc-like surface brightness distribution or there being four compact lensed images detected with the expected relative positions and peak surface brightness. The lensing nature of such events tends to be rather unambiguous. 

However, those cases that produce only two lensed images have less information for the networks to label them correctly as lensed events. Also, two distinct components within a few arcsec is a common morphology for double-lobed radio sources, which, depending on the jet-axis with respect to the observer, can have a relative peak brightness that could mimic a lensing event. In fact, during the training of the networks, the maximum separation between the multiple components in the source-plane was just 0.2 arcsec so that the background sources always had an extent that was less than the Einstein radius of the lens (we also used the simulated radio sources for generating the non-lensed visibility datasets). Therefore, double-lobed radio sources with larger separations were not presented to the networks during training. This may result in a bias towards higher FP rates for this class of radio source when the networks are applied to real ILT imaging data. 

To test this, we have generated a dataset of double-lobed radio sources in which there are two co-linear components separated within a radius of 3 arcsec from the image centre. While in the original training and testing data, the first component was always compact (to represent the radio core), here, we have randomized the separation, size and ellipticity of the two components in this dataset. In total, we generated 2000 non-lensed events; a representative sample of these is shown in Fig.~\ref{fig:fakedoubles}. We then apply the three network structures to determine whether these double-lobed radio sources are labelled as lensed or non-lensed events. 

In Fig. \ref{fig:FP_fakedoubels}, we show the FP rate of the three network structures, as a function of probability threshold. As expected, the performance of the networks to double-lobed radio sources is poor. We find that structures 1 and 3 both have a FP rate of 39.5 per cent, when the threshold is 0.99, while structure 2 has a FP rate of 6.5 per cent at the same threshold. This demonstrates that structure 2 identifies less double-lobed radio sources as lensed events, even when it is not specifically trained to recognise such samples. Also given in Fig.~\ref{fig:fakedoubles} is the predicted lensing probability and $2\sigma$ confidence interval when structure 2 is used. We see that for this representative sample, the probability is either 0, or consistent with 0 at the $2\sigma$-level for 87.5 per cent of the sample (14/16 objects). For the remaining two objects (g, j), the probability is 1.0. However, for these two cases, it is clear that the surface brightness of the double-lobed radio sources is consistent with gravitational lensing; the component with the highest flux density also has the largest angular size. Therefore, it is not surprising that the network mis-labels these two objects as likely lensed events. 

\begin{figure*}
\begin{center}

\begin{minipage}[b]{\textwidth}
\centering
\includegraphics[height=2.35cm]{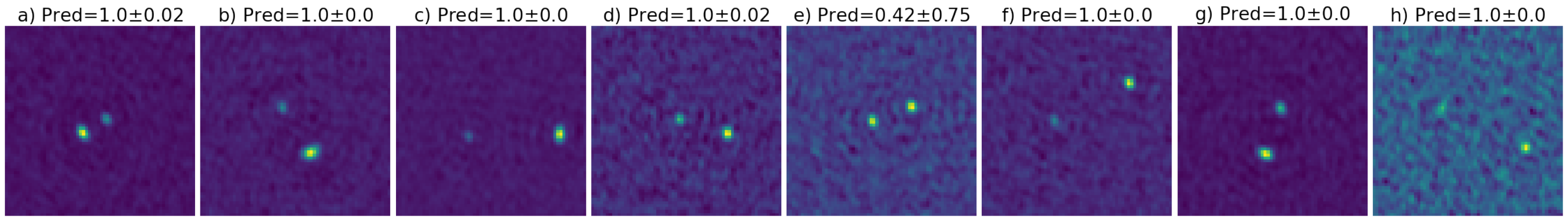}
\end{minipage}

\begin{minipage}[b]{\textwidth}
\centering
\includegraphics[height=2.35cm]{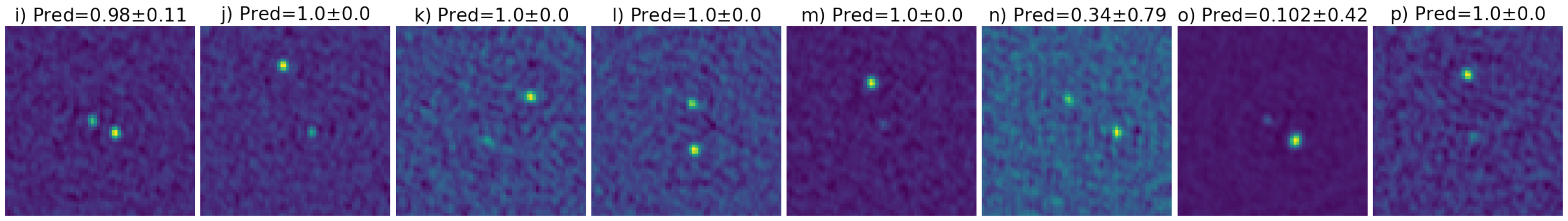}
\end{minipage}

\newpage
\caption{A representative sample of simulated gravitational lens systems with two lensed images. The average predicted probability and its uncertainty over 500 iterations for structure 2 is stated above each image. Each image contains $64\times64$~pixels and is equivalent to a sky-area of $7.68\times7.68$~arcsec$^2$.}
\label{fig:realdoubles}
\end{center}
\end{figure*}

In order to measure the performance of the three network structures when only lensed events with two images are included in the test dataset, we have generated a sample of 2000 gravitational lens systems (with realistic lens parameters) producing two distinct components that are separated by > 0.3 arcsec. We find that all three structures have a slightly poorer TP rate than before, with values of 83, 80 and 79 per cent for structures 1, 2 and 3, respectively. In Fig. \ref{fig:realdoubles}, we present a representative sample of lensed events that produce two images, with the predicted lensing probability and $2\sigma$ confidence interval, when structure 2 is used. We see that 75 per cent of the objects have a lensing probability of 1 (12/16 objects); these are all characterised with a clear surface brightness that is consistent with gravitational lensing. For 19 per cent of the objects, the probability is consistent with $>0.99$ at the $2\sigma$-level (3/16 objects), but the uncertainties are quite large, with $\sigma$ between 0.11 and 0.79. Only one object (o) has a detection probability $< 0.99$ at the $2\sigma$-level. For those cases with large uncertainties or low probabilities, the lensed images tend to have a very similar flux density or the counter-image is compact; this suggests that for these systems, the surface brightness information is insufficient to precisely label these samples as lensed events.

Our tests, using a dataset comprising double-lobed radio sources and gravitational lens systems producing two lensed images, demonstrate that structure 2 has the lowest FP rate (6.5 per cent for a threshold of 0.99) whilst having a competitive TP rate (80 per cent) to the other network structures. However, structure 2 still rates a significant number of the double-lobed radio sources as lensed events. Therefore, we re-ran our training including 2000 double-lobed radio sources, before re-testing the three networks on a dataset of 2000 previously un-seen double-lobed radio sources. We find that structures 1 and 2 returned 0 and 1 FP events (a FP rate of $<0.05$ and 0.05~per cent, respectively) and that structure 3 identified 2 FP events (a FP rate of 0.1 per cent). This highlights the importance of a comprehensive training sample in order to generate a reliable lens detection algorithm.

\subsection{Prediction uncertainties}
\label{Uncertain_Predictions}

\begin{figure}
    \centering
    \includegraphics[height=6cm]{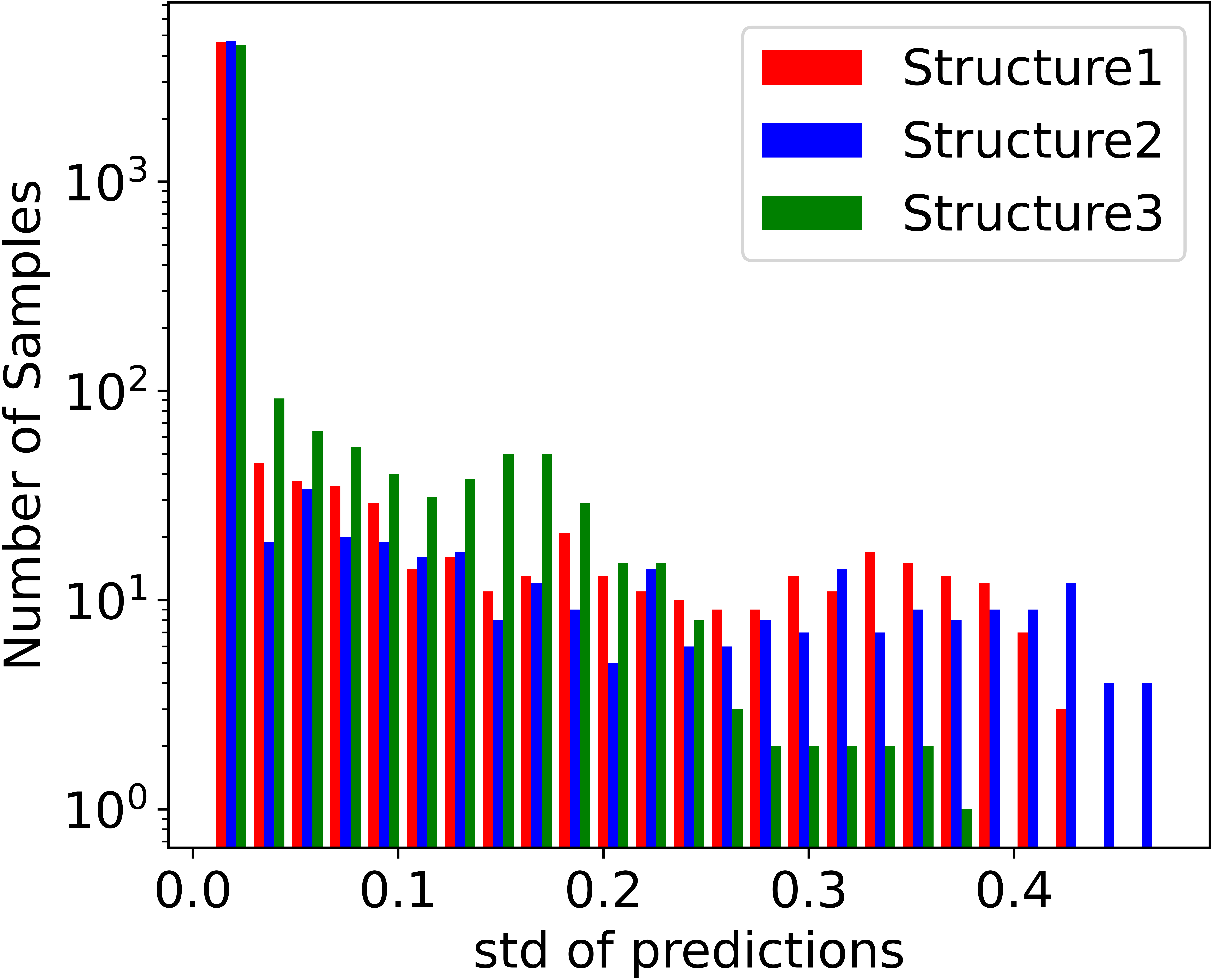}
    \caption{The standard deviation of the lensing probability for each test sample in the realistic dataset. Each sample in the test dataset has been evaluated 250 times using the dropout method.}
    \label{fig:std}
\end{figure}

\begin{figure*}
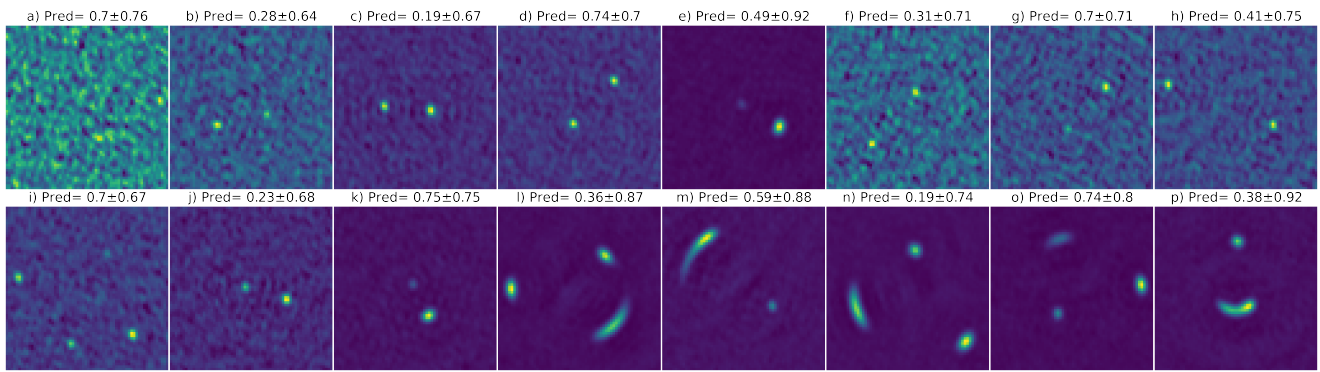

\begin{center}
\begin{minipage}[b]{\textwidth}
\centering
\includegraphics[height=2.35cm]{Fig14_1.pdf}
\end{minipage}
\newline
\begin{minipage}[b]{\textwidth}
\centering
\includegraphics[height=2.35cm]{Fig14_2.pdf}
\end{minipage}
\end{center}
\caption{A representative sample of simulated gravitational lens systems with a high uncertainty in the probability of lensing. The average predicted probability and its uncertainty over 250 iterations for structure 1 is stated above each image. The upper row (lower row) is for a lens population with a set of realistic (uniform) mass model parameters. Each image contains $64\times64$~pixels and is equivalent to a sky-area of $7.68\times7.68$~arcsec$^2$.}
\label{fig:highstd_lens}
\end{figure*}

As discussed in Section \ref{model_uncertainty}, we have used a Monte Carlo dropout technique to measure the model confidence and determine if the predicted probability of lensing is reliable or not. Human experts can then visually inspect those low reliability samples and decide if they warrant further analysis or follow-up observations. In Fig. \ref{fig:std}, we show the standard deviation of the predicted probability for each sample in the test dataset made using realistic lens model parameters, as a function of network structure. Note that the test dataset for a uniform set of lens model parameters has a similar distribution, but there are more samples with a higher standard deviation due to the poorer performance of structures 1 and 3 (see Fig.~\ref{fig:FP_TP_uniform}).  

Overall, we see that for the vast majority of cases (> 84 per cent), the standard deviation in the probability is 0. This means that the same probability is returned for each iteration during the dropout process, and therefore, the network has a confident estimate of the predicted probability. We have considered a standard deviation $> 0.3$ as non-confident cases, and have taken a closer look at those samples to understand why the network is uncertain. 

Among all the lensed samples in the test data set for a realistic lens population, there are 44 samples with a high uncertainty in which 41 cases are lenses that produce two images (see upper row in Fig.~\ref{fig:highstd_lens} for a representative sample). The reason for these samples having such a high uncertainty is likely due to the network being confused between genuine two image systems and (non-lensed) double-lobed radio sources where there is not enough information available for the network to confidently determine the predicted probability. Therefore, excluding samples with a high uncertainty will lower the TP rate, but this will also lower the FP rate for those cases of double-lobe radio sources were the network is uncertain.

In the case of the test dataset that is drawn from a population of lenses with a uniform set of mass model parameters (see lower row in Fig.~\ref{fig:highstd_lens} for a representative sample), the uncertainties are at a slightly higher level.  However, we also see sets of samples with large uncertainties (l, n, o) that have rather exotic lensing configurations, where an extended source is partially quadruply imaged, or the lens ellipticity is extremely high, resulting in an unusual image configuration. This is likely due to the limited size of the training dataset (see below for discussion). For all of the cases shown in Fig.~\ref{fig:highstd_lens}, the predicted probability of lensing is less than 0.99, and so these samples would strictly not have been selected as lens candidates even though their uncertainties are high enough for the probability to be > 0.99 (at the $2\sigma$ level).

\section{Lens detection with the ILT}
\label{results}

In this section, we use the results obtained above to design and carry out a final experiment that quantifies the reliability of our network structure for finding gravitational lenses with the ILT. We then determine the parameter space that a gravitational lens survey with the ILT would be sensitive to, in terms of the depth and angular resolution of the expected imaging data. 

\subsection{Results for the final network test} \label{finalresults}

We found from our network tests that the training and test datasets should be drawn from the same underlying lensing population, and that we must have a representative sample of non-lensed events, including both compact and double-lobed radio sources. Therefore, for our final test, we use a training dataset that has a total of 30000 samples, where 15000 are lensed events, created using a uniform distribution for the lens model parameters, and 15000 are non-lensed events, which are further divided into 11000 compact radio sources and 4000 double-lobed radio sources. This training model is then made in the same way as described in Section~\ref{method}. Again, a maximum of 450 training iterations were used.

For testing, we have used a sample comprising 17385 lensed events that were created using a uniform distribution of lens model parameters; this was done so that we can test our trained model against a wide variety of lensing configurations, including those not necessarily seen by the network. This will likely produce more conservative, but less biased results. The sample of non-lensed events includes 15385 compact radio sources and 2000 double-lobed radio sources, which ensures that the ratio of compact-to-extended radio sources is similar to that observed with the ILT \citep{Sweijen2022}. The total number of test samples was chosen so that around 15 to 25 FP events would be returned by each of the three network structures.

\begin{figure}
\begin{center}
\begin{minipage}[b]{\columnwidth}
\centering
\includegraphics[width=\columnwidth]{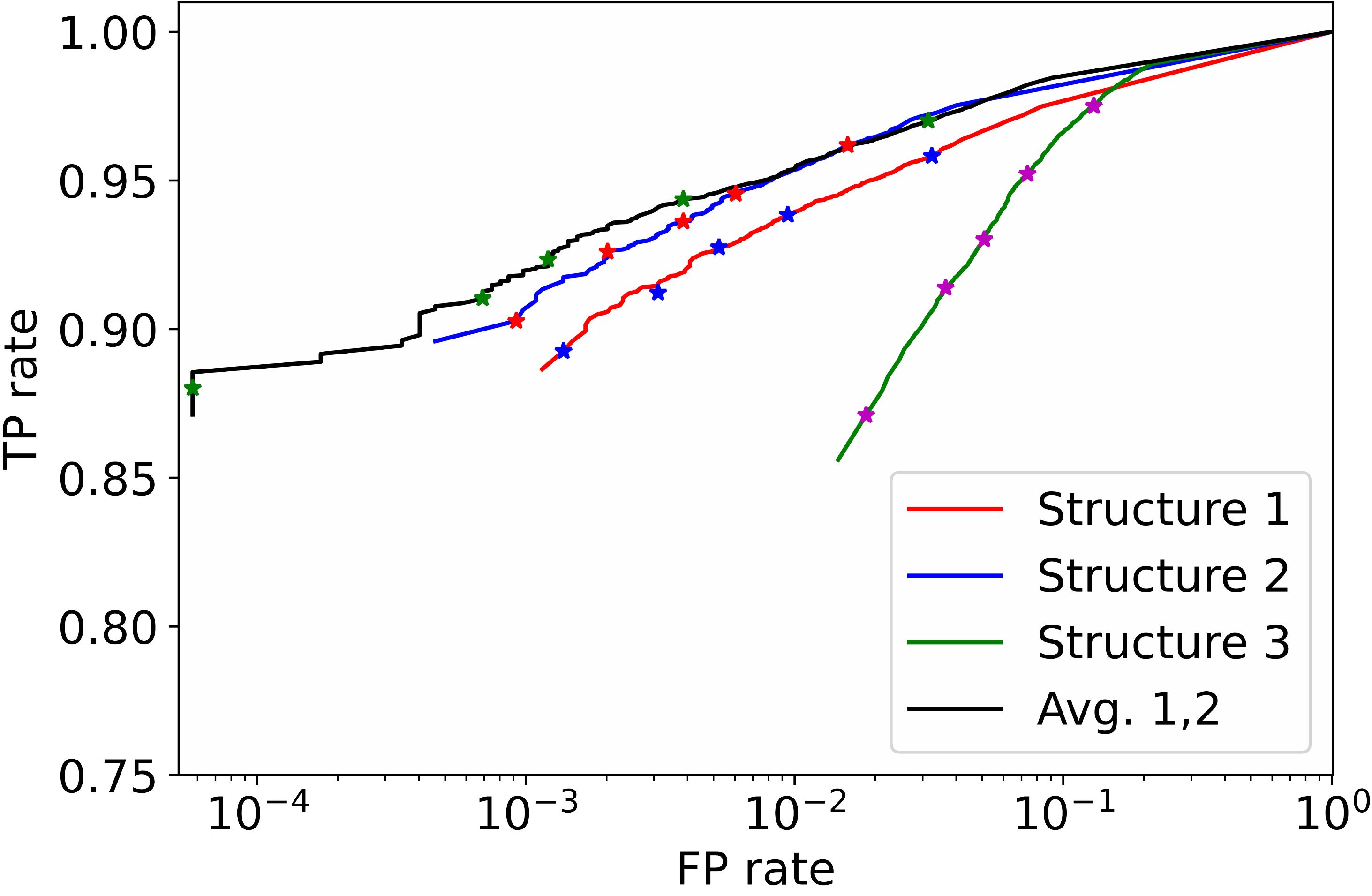}
\end{minipage}
\caption{The TP rate as a function of FP rate for the three network structures, when applied to the final dataset created using a uniform distribution of lens model parameters. Also shown are the results when network structures 1 and 2 are combined. The stars correspond to thresholds of 0.1, 0.5, 0.75, 0.9 and 0.99, from right to left.}
\label{fig:TP_LA_uniform_realistic}
\end{center}
\end{figure}

The results from this final test are shown in the ROC plots presented in Fig.~\ref{fig:TP_LA_uniform_realistic}, with the quantifiable information for a probability threshold of > 0.99 given in Table~\ref{tab:evaluation_uniform_final}. We see from Fig.~\ref{fig:TP_LA_uniform_realistic} that the performance of the three network structures in identifying lens systems is still very good, even when the network training includes double-lobed radio sources, which can introduce an additional level of confusion between lensed and non-lensed events. We see that network structures 1, 2 and 3 have TP rates of 89.3, 90.3 and 87.1 per cent, respectively. This essentially means that one in ten lenses would be missed if the network was used for lens detection as part of an ILT survey (see below for a discussion of the completeness, given the resolution and sensitivity of the ILT with respect to the flux density and angular-separation of the lensed images). Also, we see that the FP rates are 0.14, 0.09 and 1.8 per cent for structures 1, 2 and 3, respectively, which are relatively low. However, the typical probability of galaxy-scale gravitational lensing is of order $10^{-3}$, that is, one lensing event in about one thousand objects observed, which is almost identical to the FP rates of structures 1 and 2. This means that there would be between a 47 and 58 per cent chance that any lens identified by the network would be a FP event when applied to real data.

To overcome this potential issue, we have combined the results of network structures 1 and 2, which is also shown in Fig.~\ref{fig:TP_LA_uniform_realistic} and reported in Table~\ref{tab:evaluation_uniform_final}. We find that this combined network structure lowers the FP rate to 0.006 per cent (note that this is equivalent to 1 FP event in our test sample), which changes the probability to 5.7 per cent that a lens identified by the network is a FP event. This improvement in the FP rate by around an order of magnitude comes at the minimal cost of lowering the TP rate by between 1.2 to 2.2 per cent. Note that the only FP sample returned in our test dataset is a double-lobed radio source, with components that have a similar surface brightness, and a flux-ratio and component separation to double-imaged lensed events in the training dataset. Therefore, we can conclude that the imbalanced distribution of non-lensed source sizes, as shown in Fig.~\ref{fig:trainFHWM}, is not an issue for our lens detection algorithm. This is because there is sufficient information within the trained models for the network to distinguish between the lensed and non-lensed events. This is likely based on the conservation of surface brightness for the lensed events, as given by the lens equation.

\begin{table}
    \centering
\begin{tabular}{
%>{\columncolor[gray]{0.9}}
lllll}
&Structure 1&Structure 2 & Structure 3 & Avg. Str. 1+2\\
\cline{1-5}
Accuracy& 0.946 & 0.951 & 0.926 & 0.940 \\
%\cline{1-4}
{Precision}& 0.9985 & 0.9990 & 0.9792 & 0.9999 \\
%\cline{1-4}
{Recall}& 0.893 & 0.903 & 0.871 & 0.880 \\
%\cline{1-4}
{Fall out}& 0.00138 & 0.00092 & 0.01846 & 0.00006 \\
\cline{1-5}
\end{tabular}
    \caption{The evaluation criteria of the three network structures when a threshold of 0.99 is used, for the final test dataset created using a uniform distribution of lens model parameters. We also include the results when network structures 1 and 2 are combined.}
    \label{tab:evaluation_uniform_final}
\end{table}

\subsection{Parameter-space of a lens survey with the ILT} \label{parameterspace}

We now use the results from our final experiment to investigate the parameter-space, in terms of types of lenses and the brightness of the lensed images, that the network is sensitive to, given an input dataset from the ILT. For example, we expect the signal-to-noise ratio of the images to affect the reliability, as the network needs to detect multiple images to confidently predict that the object is indeed gravitationally lensed. Also, the angular resolution of the data will be an important parameter, as the lensed images need to be separable in the imaging data; not resolving all of the multiple lensed images can also lead to an inaccurate prediction of the lensing probability, particularly for exotic or high-magnification lensing events.

\begin{figure}
    \centering
    \includegraphics[width=\columnwidth]{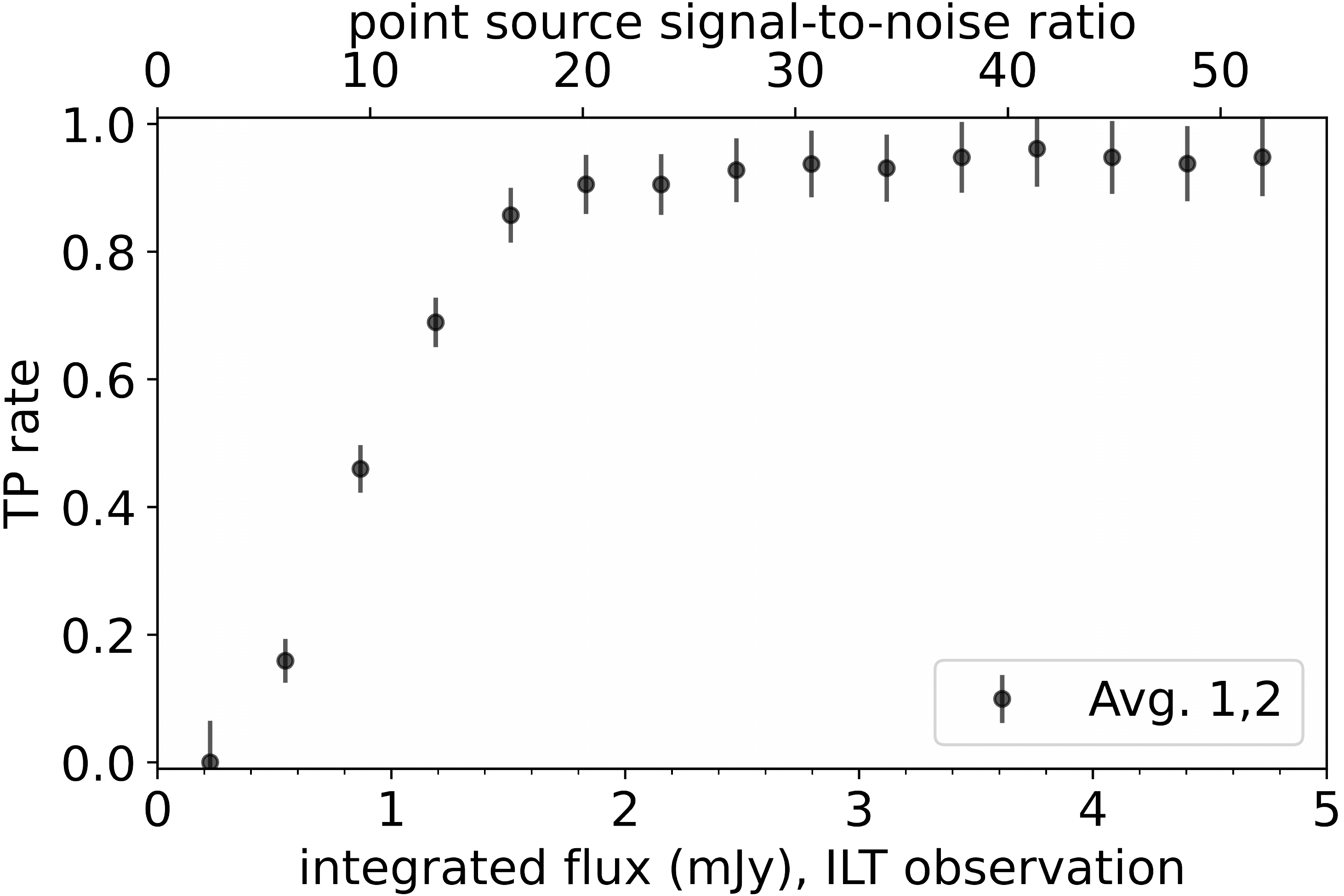}
    \caption{The TP rate as a function of the integrated flux density of the lensed radio sources from the combined results from applying network structures 1 and 2, for an ILT-like observation (rms noise 90~$\mu$Jy~beam$^{-1}$). Also shown is the point source signal-to-noise ratio for reference. Note that this is for objects with an Einstein radius $\geq0.5$~arcsec.}
    \label{fig:SB_TP}
\end{figure}

Understanding the detection rate as a function of signal-to-noise ratio is challenging to quantify, as the resulting lensed images from the same background source model can have a large variation in surface brightness distributions, depending on the lens model. For simplicity, we have analysed the detectability of a lensed event by considering the TP rate as a function of the total flux density in the model lensed images, which we show in Fig.~\ref{fig:SB_TP} (note that to separate the effect of the angular resolution, we only include the data for those systems with an Einstein radius $\geq 0.5$~arcsec; see below). This is a reasonable assumption, as the source counts of radio sources are always given as a function of flux density, as opposed to surface brightness. 

We see that for a typical ILT observation, the TP rate is of order 90 per cent for an integrated flux density of $\geq 2$~mJy, below which, the detectability of the lensed events steadily drops to just below 50 per cent at an integrated flux density of 1 mJy. We find that those lensed events with the lowest detectable surface brightness are usually in the form of a doubly- or a quadruply-imaged system with compact lensed images, which provide the least amount of information for the network to use. When the lensed events form Einstein rings and/or gravitational arcs, the integrated flux density needs to be higher for a detection to be made, and so, it is easier for the network to identify these cases at high integrated flux densities. Combined, these two effects result in the detectability of the lensed events being lower toward lower flux densities, but stops there being a very sharp cut-off. Also, we find that the cut-off is remarkably close to systems that are detected at the 15 to  $20\sigma$-level (point-source sensitivity), which in the case of doubly-imaged systems is equivalent to a flux-ratio of between 2:1 and 3:1 for the two detected lensed images (assuming detectability of emission at the $5\sigma$-level), which is fairly typical for this image configuration.

\begin{figure}
\begin{center}
\begin{minipage}[b]{\columnwidth}
\centering
\includegraphics[width=\columnwidth]{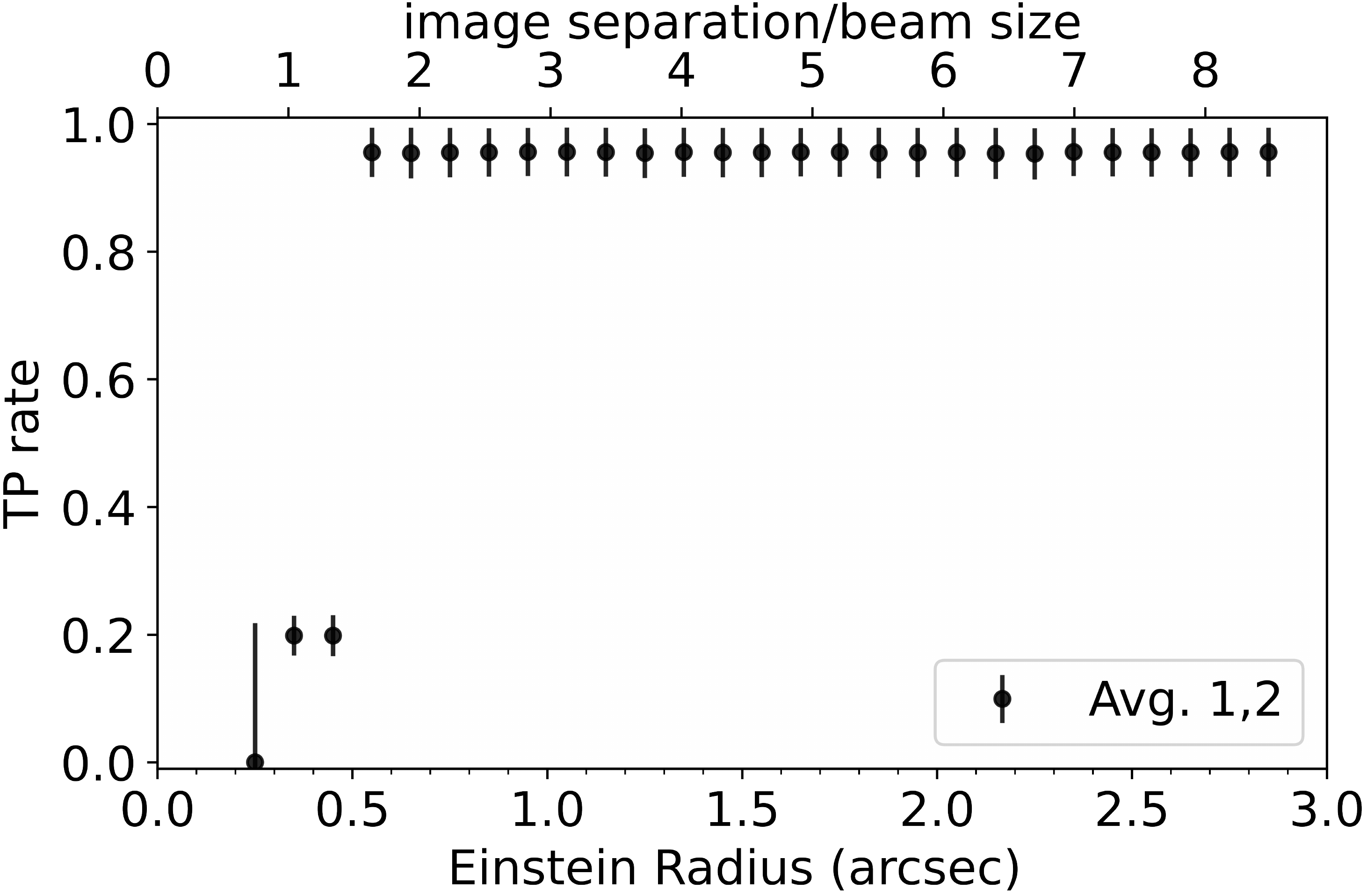}
\end{minipage}
\caption{The TP rate as a function of Einstein radius from the combined results from applying network structures 1 and 2, for an ILT-like observation (average beam size of 336 mas). Also shown is the ratio between the image separation and the average beam size for reference. Note that this is for objects with an integrated flux density $\geq 2$~mJy.}
\label{fig:TP_Re_uniform_realistic}
\end{center}
\end{figure}

The TP rate as a function of the Einstein radius is shown in Fig. \ref{fig:TP_Re_uniform_realistic} (note that to separate the effect of the sensitivity, we only include the data for those systems with an integrated flux density $\geq 2$~mJy; see above). We see that the overall TP rate is rather flat at around 95 per cent, down to an Einstein radius of about 0.5 arcsec, after which there is a sudden drop in the TP rate to about 20 per cent at Einstein radii between 0.3 and 0.5 arcsec. To some extent, this result makes sense, as an Einstein radius of 0.5 arcsec corresponds to an image separation of around 1 arcsec, which would be well resolved with the ILT (about 3 beam widths). Overall, this is an encouraging result, as it demonstrates that ILT-like imaging data is extremely sensitive to detecting gravitational lenses with Einstein radii $\geq 0.5$~arcsec, which would be sufficient to detect the majority of the gravitational lenses that are currently known (see Fig.~\ref{fig:lens_prop}).

In summary, from our simulations we find that the ILT, when included as part of LoTSS, would be most sensitive to lensed events where the integrated flux density of the lensed images is $\geq 2$~mJy and the Einstein radius of the lens model is $\geq 0.5$~arcsec. If we restrict our testing dataset to include only those lensed and non-lensed samples with these properties, we find that the overall TP rate is 95.3 per cent and the FP rate is 0.008~per cent, when network structures 1 and 2 are combined (the individual FP rates of structures 1 and 2 are also about an order of magnitude lower than reported in Table~\ref{tab:evaluation_uniform_final}). Therefore, a gravitational lens survey with the ILT, which applies these criteria and uses our algorithm for lens detection, is expected to have a completeness of 95.3 per cent and a purity of 92.2 per cent. These predictions are dependent on the lensing probability of the LoTSS source population (currently assumed to be 10$^{-3}$), which we will discuss in a future paper.

\section{Conclusions} 
\label{discussion}

In this paper, we have presented a machine learning based approach that is designed to identify galaxy-scale gravitational lenses from imaging data obtained with the ILT, taken as part of the ongoing LoTSS survey. To do this, we first simulated realistic gravitational lensing data, based on our best understanding of the radio source population at 150 MHz and the properties of galaxy-scale gravitational lenses. With these data, we have tested multiple network structures to determine whether a single or combined network produces the best results. We find that our ability to correctly identify lensed features is highly dependent on the training dataset used, which given the large parameter-space available for both the lens and source models (and their combination), represents the most challenging aspect of our method. However, by using a combined network strategy and limiting the parameter-space of the models to include only those gravitational lenses with an Einstein radius $\geq0.5$~arcsec and a total flux density of $\geq2$~mJy for the lensed images, we find that our lens detection strategy can recover 95.3 per cent of the simulated lens systems, with a FP rate of just 0.008 per cent (equivalent to a sample purity of 92.2 per cent for a lensing optical depth of 10$^{-3}$). 

Given the angular resolution and sensitivity of the imaging data to be taken during LoTSS, we conclude that the ILT has the requirements to be a gravitational lens finding machine, and that deep learning techniques provide an efficient route to discovering new gravitational lenses with interferometric imaging data. We find that data, with a similar resolution and noise properties to the ILT, can be used to find gravitational lenses with lensed image separations that are $\geq3$ times the synthesized beam width, when the lensed images are detected at the $\geq20\sigma$-level. This suggests that the SKA-MID, with baselines of up to 150 km and an angular resolution of 0.3 arcsec should also be excellent at finding new gravitational lens systems in the future. As a next step, we will carry out a similar analysis with dedicated SKA-MID gravitational lensing simulations to confirm this, and to optimise lens searches with this next generation instrument.

Our current set of simulations, although realistic, is likely the limiting factor in our analysis, and will be further improved in the future. First, we will increase the range of lens models probed to include elliptical power-law mass distributions, as opposed to focusing on only isothermal cases. We have also included a somewhat simple parameterisation of the background radio source structure, which was computationally easy to implement, but was also driven by our lack of knowledge of the low frequency radio source population. As LoTSS progresses, and the level of information about the structure of radio sources on 0.3 arcsec-scales improves, we will revisit our simulations to refine our network structure with a better model for the radio source population. This will also involve training on a larger number of lens and source model combinations to better sample the available parameter-space. Our simulations have also assumed perfectly calibrated data, and although small-scale calibration errors will exist in the real imaging data, their impact on our ability to find gravitational lenses is not clear, and will be addressed in a follow-up paper. Finally, we intend to include the frequency information (radio spectral index), given the wide bandwidth of the ILT data, to further separate lensed and non-lensed events.

Overall, our results are encouraging for finding new galaxy-scale gravitational lenses with the ILT, and in the short term we will apply our best trained model to the first tranche of data from LoTSS that includes the ILT baselines (e.g. \citealt{Morabito2021,Sweijen2022}) and has detected over 2500 radio sources at an angular resolution of around 350 mas. Given the expected probability for galaxy-scale gravitational lensing, there should be between 2 and 4 systems within such a dataset, which we are now focusing on finding. As more LoTSS observing fields are processed with the ILT data included, we expect the first new discoveries of gravitational lensing with LOFAR to be made. These new systems will be used to test and further refine our network structure so that we are in the best possible position to analyse the data for the up to 15 million radio sources that will be observed during LoTSS.

\section*{Acknowledgements}
We thank Kerstin Bunte and the members of the LOFAR Gravitational Lensing working group within the LOFAR Surveys KSP, and in particular, Philippa  Hartley and Hannah Stacey for their useful discussions. This paper is based on research performed within in the DSSC Doctoral Training Programme, co-funded through a Marie Skłodowska-Curie COFUND (DSSC 754315). JPM acknowledges support from the Netherlands Organization for Scientific Research (NWO) (Project No. 629.001.023) and the Chinese Academy of Sciences (CAS) (Project No. 114A11KYSB20170054). LOFAR \citep{vanHaarlem2013} is the Low Frequency Array designed and constructed by ASTRON. It has observing, data processing, and data storage facilities in several countries, which are owned by various parties (each with their own funding sources), and that are collectively operated by the ILT foundation under a joint scientific policy. The ILT resources have benefited from the following recent major funding sources: CNRS-INSU, Observatoire de Paris and Universit\'e d'Orl\'eans, France; BMBF, MIWF-NRW, MPG, Germany; Science Foundation Ireland (SFI), Department of Business, Enterprise and Innovation (DBEI), Ireland; NWO, The Netherlands; The Science and Technology Facilities Council, UK; Ministry of Science and Higher Education, Poland; The Istituto Nazionale di Astrofisica (INAF), Italy. The J\"ulich LOFAR Long Term Archive and the German LOFAR network are both coordinated and operated by the J\"ulich Supercomputing Centre (JSC), and computing resources on the supercomputer JUWELS at JSC were provided by the Gauss Centre for Supercomputing e.V. (grant CHTB00) through the John von Neumann Institute for Computing (NIC).

\section*{Data Availability}

Upon reasonable request, the underlying data used for this article will be shared by the corresponding author.


\bibliographystyle{mnras}
\bibliography{example} % if your bibtex file is called example.bib

\bsp	% typesetting comment
\label{lastpage}
\end{document}